\documentclass[12pt,dvips]{article}

\usepackage{fullpage,epsfig,fancyheadings}

\usepackage[psamsfonts]{euscript}

\usepackage{epic}

\usepackage{amstex}

\numberwithin{equation}{section} 

\begin{document}
\tolerance=10000
\hfuzz=5pt
\vspace*{-.6in}
\thispagestyle{empty}
\begin{flushright}
UW/PT 96-03\\
CALT-68-2014\\
IFUM/FT 517
\end{flushright}

\vspace{.5in}
{\Large
\begin{center}
Confinement:  Understanding the Relation Between the Wilson Loop and Dual
Theories of Long Distance Yang Mills Theory.
\end{center}}
\vspace{.4in}

\begin{center}
M. Baker\footnote{Work supported in part by the U.S. Dept. of
Energy under Grant No. DOE/ER/40614.}\\
{\it University of Washington, Seattle, WA 98105, USA}\\
\medskip
J.S. Ball\\
{\it University of Utah, Salt Lake City, UT  84112, USA}\\
\medskip
N. Brambilla\\
\medskip
G.M. Prosperi\\
{\it Dipartimento di Fisica dell'Universita-Milano}\\
{\it Istituto Nazionale di Fisica Nucleare, Sezione}\\
{\it di Milano - Via Celoria 16, 20133 Milano, Italy}\\
\medskip
F. Zachariasen\footnote{Work supported in part by the U.S. Dept. of
Energy under Grant No. DE-FG03-92-ER40701.}\\
{\it California Institute of Technology, Pasadena, CA 91125, USA}
\newpage

\end{center}
\vspace{1in}

\begin{center}
\textbf{Abstract}
\end{center}
\begin{quotation}
In this paper we express the velocity dependent, spin dependent heavy quark
potential $V_{q\bar q}$ in QCD in terms of a Wilson Loop $W(\Gamma)$ determined
by pure Yang Mills theory.  
  We use an effective dual theory
of long-distance Yang Mills theory to calculate $W(\Gamma)$ for large loops;
i.e.  for loops of size $R > R_{FT}$.  ($R_{FT}$ is
the flux tube radius, fixed by the value of the Higgs (monopole) mass of the 
dual theory, which is a concrete realization of the
Mandelstam 't Hooft dual superconductor mechanism of confinement).
  We replace $W(\Gamma)$ by $W_{eff}(\Gamma)$,
given by a functional integral over the dual variables, which for $R > R_{FT}$
can be evaluated by a semiclassical expansion, since the dual theory is weakly
coupled at these distances.  The classical approximation
gives the leading contribution to $W_{eff}(\Gamma)$ and yields a velocity
dependent heavy quark potential  which for large $R$ becomes linear in $R$,
and which for small $R$ approaches lowest order
perturbative QCD. 
This latter fact  means that these results should remain applicable 
 down to distances  where radiative corrections giving rise to 
a running coupling constant become important. 
The spin dependence  of the potential  
reflects the vector coupling of the quarks 
at long range as well as at short range. The methods developed
here should be applicable to any realization of the dual superconductor
mechanism. They give an expression determining $W_{eff}(\Gamma)$ independent
of the classical approximation, but  semi classical corrections due to 
fluctuations of the flux tube are not worked out in this paper.
Taking these into account should lead to an effective string theory free from
the conformal anomaly.

\end{quotation}
\vfil

\newpage

\pagenumbering{arabic} 

\baselineskip = 24pt

\section{Introduction}

In this paper we give expressions for the heavy quark potential in QCD using an
effective dual theory of long distance Yang--Mills theory.
 This work  goes beyond a previous treatment$^{[1]}$ where the quark motion
 was treated semi--classically and where the dual theory was considered
 only at the classical level, and provides an independent approach 
 to the problem  of the heavy quark potential.

In Section two we give the formulae for the heavy quark spin dependent
velocity dependent potential $V_{q\bar q}$ obtained  in refs.
$^{[2,3,4]}$ in terms of a Wilson loop $W(\Gamma)$.  This expression
extends previous work of Eichten and Feinberg,$^{[5]}$ Peskin$^{[6]}$,
 Gromes 
and others$^{[6]}$ to include the velocity dependent spin independent
part of the potential.  The problem of the heavy quark potential is
then reduced to the problem of calculating $W(\Gamma)$ in pure
Yang--Mills theory.  All momenta, spins, masses and quantum mechanical
properties of the quarks appear explicitly in the formulae$^{[2]}$
 relating $V_{q\bar q}$ to $W(\Gamma)$.  The size of the loop $\Gamma$
fixed by the classical trajectories of the moving quark-antiquark pair
provides a length scale $R$ (the quark-antiquark separation) and we use
the dual theory to evaluate $W(\Gamma)$ for $R > R_{FT}$, the radius of
the flux tube that forms between the moving quark antiquark pair.

In Section three we describe the dual theory 
 and then show how to  calculate the Wilson loops of Yang Mills theory
 at long distances (large loops).  This is done by
replacing $W(\Gamma)$ by $W_{eff} (\Gamma)$, a functional integral over
dual potentials $C_\mu$ which are the fundamental variables of the dual
theory.  
We then obtain the spin independent part of the heavy quark potential
 directly in terms of $W_{eff}(\Gamma)$.
 Finally we discuss the relation of the
dual theory to recent work$^{[7,8]}$ on the use of electric-magnetic
duality to determine the long distance behavior of certain
supersymmetric non-Abelian gauge theories.

In Section four we give explicit expressions for the spin dependent part
of the heavy quark potential in terms of quantities determined by
the dual theory. Since the theory is weakly coupled at large distances, $W_{eff}
(\Gamma)$ and hence $V_{q \bar{q}}$ can be evaluated  by a semiclassical
expansion.

In Section five we calculate $W_{eff} (\Gamma)$ in the classical
approximation to the dual theory.  We show how these results yield the
dual superconducting picture of confinement and discuss their relation
to the ``modified area law'' model for $W(\Gamma)$, proposed in
ref.$^{[9]}$.  Finally we remark how recent progress on quantization
around classical vortex solutions$^{[10]}$ may be useful for calculating
corrections to
$W_{eff}(\Gamma)$   accounting  for  fluctuations of the length of the
flux tube.

In Section six we use the results of Section four and the classical solution to
the dual theory to obtain the spin dependent part of the heavy quark potential.
 This calculation gives a contribution to $V_{q\bar{q}}$
not obtained previously$^{[11]}$  and yields a 
simplified expression for the spin orbit potential which reflects
the vector nature of both the short range force and the confinement force.

In the conclusion we point out that the results presented here should be
regarded more as consequences of the dual superconductor picture in general
rather than of our particular realization of it.$^{[12]}$

\section{The Heavy Quark Potential in QCD}

To obtain the heavy quark potential $V_{q\bar q}$ we$^{[2]}$ make
 a Foldy Wouthuysen
transformation  on the quark-antiquark Green's function and
show that the result can be
written as a Feynman path integral over particle and anti-particle coordinates
and momenta of a Lagrangian depending only upon the spin, coordinates, and
momenta of the quark and antiquark.  Separating off the kinetic terms from
this Lagrangian one can identify what remains as the heavy quark potential
$V_{q\bar q}$.  (Closed loops of light quark pairs
 and annihilation contributions
were not included.)

The terms in $V_{q\bar q}$ of order (quark mass)$^{-2}$ are of two types;
velocity dependent $V_{VD}$ and spin dependent $V_{SD}$.  The full potential
$V_{q\bar q}$ is then
\begin{equation}
V_{q \bar q} = V_0 (R) + V_{VD} + V_{SD},
\end{equation}
where $V_0(R)$ is the static potential.  These potentials are all expressed in
terms of a Wilson Loop $W(\Gamma)$ determined by pure Yang-Mills theory,
given by
\begin{equation}
W (\Gamma) = {\int {\cal D} Ae^{iS_{YM}(A)} tr P\exp (-ie \oint_\Gamma dx^\mu
A_\mu (x))\over \int {\cal D} Ae^{iS_{YM} (A)}}.
\end{equation}

The closed loop $\Gamma$ is defined by quark (anti-quark) trajectories $\vec
z_1 (t) (\vec z_2 (t))$ running from $\vec y_1$ to $\vec x_1 (\vec x_2$ to
$\vec y_2$) as $t$ varies from the initial time $t_i$ to the final time $t_f$.
The quark (anti-quark) trajectories $\vec z_1 (t) (\vec z_2 (t)$) define world
lines $\Gamma_1 (\Gamma_2)$ running from $t_i$ to $t_f (t_f$ to $t_i$).   The
world lines $\Gamma_1$ and $\Gamma_2$, along with two straight lines at fixed
time connecting $\vec y_1$ to $\vec y_2$ and $\vec x_1$ to $\vec x_2$, then
make up the contour $\Gamma$ (see Fig.1).
   As usual $A_\mu (x) = {1\over 2}
\lambda_a A_\mu^a (x)$, $tr$ means the trace over color indices,  $P$
prescribes
the ordering of the color matrices according to the direction fixed on the loop
and $S_{YM}(A)$ is the Yang--Mills action including a gauge fixing term.  We
have denoted the Yang--Mills coupling constant by e, i.e.,
\begin{equation}
\alpha_s = {e^2\over 4\pi}.
\end{equation}

The spin independent part of the potential, $V_0 + V_{VD}$, is obtained from
the zero order and the quadratic
terms in the expansion of $i\log W(\Gamma)$ for small velocities ${\dot{\vec
z}}_1 (t)$ and ${\dot{\vec z}}_2 (t)$.  This expansion has the form:
\begin{equation}
i \log W(\Gamma) = \int_{t_{i}}^{t_{f}} dt \left(
V_0 (R(t)) + \sum_{i,j=1}^2 \sum_{k,\ell =1}^3 \dot z_i^k (t) V_{ij}^{k\ell}
(\vec R(t)) \dot z^\ell_j (t)\right).
\label{lw}
\end{equation}
where $\vec R(t) = \vec z_1(t) - \vec z_2 (t)$, and
\begin{equation}
V_{V D} = \sum_{i,j=1}^2 \sum_{k,\ell=1}^3 \dot z_i^k (t) V_{ij}^{k\ell} (\vec
R(t)) \dot z^\ell_j (t).
\label{vvd}
\end{equation}
($i \log W(\Gamma)$ has an expansion of the form (\ref{lw}) only to second
order in
the velocities.) The expression  (\ref{vvd}) for  $V_{VD}$ follows from the
same
argument used to identify $V_0(R)$ as the velocity independent term in
the expansion (\ref{lw}).  We can write eq. (\ref{lw}) in the form
\begin{equation}
i \log W (\Gamma) = - \int_{t_{i}}^{t_{f}} dt L_I (z_1, z_2, \dot z_1, \dot
z_2),
\end{equation}
where
\begin{equation}
- L_I = V_0 (R) + V_{VD}
\end{equation}
is an effective interaction Lagrangian for classical particles moving along
trajectories $\vec z_1(t)$ and $\vec z_2(t)$ with gauge couplings $e (-e)$ and
we can then interpret $i\log W(\Gamma)$ as an effective action describing the
motion of classical particles after elimination of the Yang--Mills field.

The spin dependent potential $V_{SD}$ contains structures for each quark
analogous to those obtained by making a Foldy Wouthuysen transformation on the
Dirac equation in an external field $F_{\mu\nu}^{EXT}$, along with an additional
term $V_{SS}$ having the structure of a spin-spin interaction.  We can then
write
\begin{equation}
V_{SD} = V_{LS}^{MAG} + V_{Thomas} + V_{Darwin} + V_{SS},
\label{vsd}
\end{equation}
using a notation which indicates the physical significance of the individual
terms (MAG denotes magnetic).The first two terms in eq. (\ref{vsd})
can be obtained by making the
replacement
\begin{equation}
F_{\mu\nu}^{EXT} (x) \rightarrow \langle\!\langle F_{\mu\nu} (x)
\rangle\!\rangle ,
\label{fex}
\end{equation}
in the corresponding expression for the interaction of a Dirac particle in an
external field, where
\begin{equation}
\langle\langle f(A)\rangle\rangle \equiv {\int {\cal D} Ae^{iS_{YM} (A)} tr P
\{\exp [-ie \oint_{\Gamma} dx^\mu A_\mu (x)]f(A)\}\over\int {\cal D} Ae^{iS_{YM}
(A)} tr P\exp [ -ie \oint_{\Gamma} dx^\mu A_\mu (x)]},
\end{equation}
and
\begin{equation}
F_{\mu\nu} = \partial_\mu A_\nu - \partial_\nu A_\mu - ie [A_\mu, A_\nu],
\end{equation}
i.e. $\langle\!\langle F_{\mu\nu} (x) \rangle\!\rangle$ is the expectation of
the Yang--Mills field tensor in the presence of a quark and anti--quark moving
along classical trajectories $\vec z_1 (t)$ and $\vec z_2 (t)$ respectively.

The explicit expressions for $V_{LS}^{MAG}$ and $V_{Thomas}$ obtained 
in ref.$^{[2]}$
are\footnote{Here and in the following $\int_{\Gamma_j} dx^{\mu}
 f_{\mu}(x) \equiv (-1)^{j+1} \int_{t_i}^{t_f} dt ( f_0(z_j) -\vec{\dot{z}}_j
\cdot \vec f (z_j))$, where $z_j=(t,\vec{z}_j(t))$.
  The factor $(-1)^{j+1}$  accounts for the fact that
  world line $\Gamma_2 $  runs from $t_f$ to $t_i$.
We also use the notation $z_j^{\prime}=(t^{\prime},\vec{z}_j(t^{\prime}))$.} 
\begin{equation}
\int dt V_{LS}^{MAG} =  \sum_{j=1}^2 {e\over m_j} \int_{\Gamma_{j}}
dx^\sigma S_j^\ell \langle\!\langle \hat F_{\ell\sigma} (x)\rangle\!\rangle,
\label{vls}
\end{equation}
and
\begin{equation}
\int dt V_{Thomas} = -\sum_{j=1}^2 {e\over 2m_j^2} \int_{\Gamma_{j}}
dx^\mu S_j^\ell \epsilon^{\ell kr}p_j^k \langle\!\langle F_{\mu r}
(x)\rangle\!\rangle ,
\label{vt}
\end{equation}
where
\begin{equation}
\hat F_{\mu\nu} = -{1\over 2} \varepsilon_{\mu\nu\rho\sigma} F^{\rho\sigma},
\end{equation}
$\vec{S}_j$ is the
 spin matrix, and $m_j$ is the mass of the $j$th quark.
Because the expression 
 for $V_{Thomas}$ contains an explicit factor of $1/m_j^2$,
 the integral over the trajectory of the $j$th quark 
$\int_{\Gamma_{j}} dx^\mu \langle\!\langle F_{\mu r} (x) \rangle\!\rangle$
 can be replaced by $ (-1)^{j+1}
\int_{t_i}^{t_f} dt$   $\langle\!\langle F_{0 r}
( z_j)\rangle\!\rangle$ evaluated for static quarks.  This gives
the usual expression for $V_{Thomas}$ in terms of the derivative of the central
potential (see Section four).  The expression for $V_{LS}^{MAG}$ on the other
hand
contains only a single power of $1\over m_j $ and $\int_{\Gamma_{j}} dx^\sigma
\langle\!\langle \hat F_{\ell\sigma} (x) \rangle\!\rangle$ must be evaluated to
first order in the quark velocities.  There results the usual
magnetic interaction of
the spin of the $j$th quark with the expectation value
$\langle\!\langle \hat F_{\ell\sigma} (z_j)
 \rangle\!\rangle$.

The expression for  $V_{Darwin}$ 
is 
\begin{equation}
\int dt V_{Darwin} = -\sum_{j = 1}^2 {e\over 8m_j^2} \int_{\Gamma_{j}}
dx^\mu \langle\!\langle D^\nu F_{\nu\mu} (x) \rangle\!\rangle,
\label{vd}
\end{equation}
where
\begin{equation}
D^\nu F_{\nu\mu} = \partial^\nu F_{\nu\mu} - i e [A^\nu, F_{\nu\mu}].
\end{equation}
Again because of the explicit factor of
${1\over m_j^2}$, the integral over the trajectory $\Gamma_j$ of the $j$th
quark is  evaluated for static quarks.

The final term $V_{SS}$ in eq. (\ref{vsd}) is given by
\[
\int V_{SS} dt = - {1\over 2} \sum_{j,j'=1}^2 {ie^2\over m_j m_{j'}} T_s
\int_{\Gamma_{j}} dx^\mu \int_{\Gamma_{j'}} dx^{\prime\sigma} S_j^\ell
S_{j'}^k\]
\begin{equation}
(\langle\!\langle \hat F_{\ell\mu} (x) \hat F_{k\sigma} (x') \rangle\!\rangle -
\langle\!\langle \hat F_{\ell\mu} (x) \rangle\!\rangle \langle\!\langle \hat
F_{k\sigma} (x') \rangle\!\rangle),
\label{vss}
\end{equation}
where $T_s$ is the spin time ordering operator along the paths $\Gamma_1$ and
$\Gamma_2$, and the averages are  evaluated for static quarks.  The terms
$j
\neq j'$ in eq. (\ref{vss}) give a spin-spin interaction proportional to $1/m_1
m_2$
while the terms $j = j'$ in eq. (\ref{vss}) give a spin independent term
proportional to $({1\over m_1^2} + {1\over m_2^2})$. The spin ordering is
relevant only for these latter terms.

We have thus obtained the  explicit expression (2.8)  for the spin dependent 
 potential 
as a sum of terms depending upon the quark and antiquark spins, masses and
momenta with coefficients which are expectation values $\langle\!\langle
{}~\rangle\!\rangle$ of operators computed in Yang--Mills theory in presence of
classical sources generated by the moving quark-antiquark pair.  We now show
that these expectation values can be obtained as functional derivatives of $i
\log W(\Gamma)$ with respect to the path, i.e., with respect to the
trajectories $\vec z_1 (t)$ or $\vec z_2 (t)$.  For example consider the change
in $W(\Gamma)$ induced by letting
\[
\vec z_1 (t) \rightarrow \vec z_1 (t) + \delta \vec z_1 (t), ~~{\rm where}~~
\delta \vec z_1  (t_i) = \delta \vec z_1 (t_f) = 0.\]
Then from the definitions (2.2) and (2.10), it follows that
\begin{equation}
\delta i \log W(\Gamma) =-{ e}
 \int_{t_{i}}^{t_{f}} {\delta S^{\mu\nu} (z_1) \over 2}
\langle\!\langle F_{\mu\nu} (z_1) \rangle\!\rangle,
\label{e18}
\end{equation}
where
\begin{equation}
\delta S^{\mu\nu} (z_1) =  (dz_1^\mu \delta z_1^\nu - dz_1^\nu
\delta z_1^\mu).
\label{smn}
\end{equation}
\noindent
Eq. (\ref{e18}) then gives
\begin{equation}
-{e}
\langle\!\langle F_{\mu\nu} (z_1) \rangle\!\rangle = {\delta i \log
W(\Gamma)\over \delta S^{\mu\nu} (z_1)} ,
\label{e20}
\end{equation}
and similarly one can get
\[
{e}
\langle\!\langle F_{\mu\nu} (z_2) \rangle\!\rangle = {\delta i \log
W(\Gamma)\over \delta S^{\mu\nu} (z_2)} .
\]

\noindent
Varying the path $\vec z_2 (t)$ in eq. (\ref{e20}) gives
\begin{equation}
e^2 \left(\langle\!\langle F_{\mu\nu} (z_1) F_{\rho\sigma} (z_2)
\rangle\!\rangle - \langle\!\langle F_{\mu\nu} (z_1)
\rangle\!\rangle \langle\!\langle F_{\rho\sigma} (z_2)\rangle\!\rangle \right)
= i e {\delta\over \delta S^{\rho\sigma}
(z_2)} \langle\!\langle F_{\mu\nu} (z_1) \rangle\!\rangle.
\label{e21}
\end{equation}

The first and second variational derivatives of $W(\Gamma)$ 
 then  determine the
expectation values of  $F_{\mu\nu}$
needed to evaluate $V_{SD}$.
Furthermore, we show in
an appendix that 
$\langle \langle D^{\nu} F_{\nu\mu} (x) \rangle\rangle$ appearing
in $V_{Darwin}$ can also be expressed in terms of variational derivatives of
$W(\Gamma)$.  The Wilson loop $W(\Gamma)$ which is determined by pure
Yang--Mills theory then fixes the complete heavy quark potential $V_{q\bar q}$.
 Thus, up to order $\left({1\over
quark \,  mass}\right)^2$ the dynamics of a quark anti-quark pair in QCD
is completely fixed by the dynamics of Yang--Mills theory.  The properties of
the quark spins, masses, etc., appear only as given kinematic factors in the
terms defining the heavy quark potential.

The result (2.8) for $V_{SD}$ is a
consequence of the vector nature of the QCD interaction and contains precisely
the same dependence upon the quarks spins, masses, and momenta as in QED.  For
example in eqs. (\ref{vls}) and (\ref{vt}) there is the usual vector coupling
of quarks
to $\langle\!\langle F_{\mu\nu} (x) \rangle\!\rangle$.  The long (short) range
part of $V_{SD}$ is determined by the behavior of this field at long (short)
distances.  Both have the same vector coupling.

This expansion as it stands is applicable only to calculating the potential
between heavy quarks.  The essence of the constituent quark model is that the
same potential can also be used to calculate the energy levels of mesons
containing light quarks with constituent masses fixed by hadron spectroscopy.
The assumption is that the principal effect of the light quark dynamics
can be accounted for by giving the light quarks effective masses which become
the parameters of the constituent quark model.

Finally we  note the following "modified area law" proposed in ref.$^{[9]}$:
   $i \log W(\Gamma)$ is written as the sum of a short range (SR)
contribution and a long range (LR) one:
\begin{equation}
i \log W(\Gamma) = i \log W^{SR} (\Gamma) + i \log W^{LR} (\Gamma),
\end{equation}
with $i \log W^{SR} (\Gamma)$  given by ordinary perturbation
theory and 
\begin{equation}
i \log W^{LR} (\Gamma) = \sigma S_{min},
\end{equation}
where $S_{min}$ is the minimal surface enclosed by the loop $\Gamma$ and
$\sigma$ is the string tension.  We will see in Section five the relation of
this ansatz to the predictions of the dual theory.

\section{The Dual Description of Long Distance Yang-Mills Theory}

The dual theory is an effective theory of long distance Yang--Mills theory
described by a Lagrangian density ${\cal L}_{eff}$ in which the fundamental
variables are an octet of dual potentials $C_\mu$ coupled minimally to three
octets of scalar Higgs fields $B_i$ carrying magnetic color charge.  (The gauge
coupling constant of dual theory $g = {2\pi\over e}$ where $e$ is the
Yang--Mills coupling constant.)  The monopole fields $B_i$ develop
non-vanishing vacuum expectation values $B_{0i}$ (monopole condensation) which
give rise to massive $C_\mu$ and consequently to a dual Meissner effect.  Dual
potentials couple to electric color charge like ordinary potentials couple to
monopoles.  The potentials 
$C_\mu$ thus couple to a quark anti-quark pair via a Dirac
string connecting the pair.  The dual Meissner effect prevents the electric
color flux from spreading out as the distance $R$ between the quark anti-quark
pair increases. As a result
 a linear potential develops which confines the quarks in
hadrons.  The dual theory then provides a concrete realization of the
Mandelstam 't Hooft$^{[13]}$ dual superconductor picture of
confinement.

Because the quanta
of the potentials $C_\mu$ are massive, the dual theory is weakly coupled at
distances $R> {1\over M}$ ($M$ being either the mass of the dual gluon or of
the monopole field)
 and a semi-classical expansion can be used to calculate the
heavy quark potential at those distances.
  The classical approximation gives the leading
contribution to functional integrals defined by ${\cal L}_{eff}$, in contrast
to the functional integrals of Yang--Mills theory where no single configuration
of gauge potentials dominates $W(\Gamma)$.  The duality 
assumption that the long
distance physics of Yang--Mills theory depending upon strongly coupled gauge
potentials $A_\mu$ is the same as the long distance physics of the dual theory
describing the interactions of weakly coupled dual potentials $C_\mu$ and
monopole fields $B_i$ forms the basis of the work of this paper.

Before writing down the explicit form of  ${\cal L}_{eff}$,
we first show how to calculate $W(\Gamma)$ for Abelian Gauge theory using the
dual description of electrodynamics$^{[14]}$, which describes the same
physics as the original description at all distances.  We consider a pair of
particles with charges $e (-e)$ moving
along trajectories $\vec z_1(t) (\vec z_2 (t))$ in a relativistic medium having
dielectric constant $\epsilon$.  The current density $j^\mu
(x)$ then has the form
\begin{equation}
j^\mu (x) = e \oint_\Gamma dz^\mu \delta (x - z),
\label{j}
\end{equation}
where $\Gamma$ is the world line described in fig.1.
In the usual $A_\mu$ (electric) description this system is described by a
Lagrangian
\begin{equation}
{\cal L}_A(j) = - {\epsilon\over 4} (\partial_\alpha A_\beta - \partial_\beta
A_\alpha)^2 - j^\alpha A_\alpha,
\label{la}
\end{equation}
from which one obtains the usual Maxwell equations
\begin{equation}
\partial^\alpha \epsilon (\partial_\alpha A_\beta - \partial_\beta A_\alpha) =
j_\beta .
\label{hma}
\end{equation}
If the (wave number dependent) dielectric constant $\epsilon \rightarrow 0$ at
long distances, then we see from eq. (3.3) that $A_\mu$ is strongly
coupled at
long distances (anti-screening).  From (\ref{j}) and (\ref{la}) we have
\begin{equation}
\int dx {\cal L}_A (j) = - \int dx {\epsilon (\partial_\mu A_\nu - \partial_\nu
A_\mu)^2\over 4} - e \oint_\Gamma dz^\mu A_\mu (z).
\end{equation}
The functional integral defining $W(\Gamma)$ in Abelian gauge theory
\begin{equation}
W(\Gamma) = {\int {\cal D} A_\mu e^{i\int dx [{\cal L}_A (j) + {\cal
L}_{GF}]}\over \int {\cal D} A_\mu e^{i\int dx [{\cal L}_A (j = 0) + {\cal
L}_{GF}]}},
\end{equation}
where ${\cal L}_{GF}$ is a gauge fixing term, is gaussian and has the value
\begin{equation}
W(\Gamma) = e^{{ie^{2}\over 2}\int_\Gamma dx^\mu \int_{\Gamma'} dx^{\prime\nu}
{D_{\mu\nu}(x - x')\over \epsilon}},
\label{wg}
\end{equation}
where $D_{\mu\nu}$ is the free photon propagator and where
self energies have been subtracted. Because of current
conservation the result (\ref{wg}) is independent of the choice of gauge.
Letting $\epsilon = 1$ and expanding $i \log W(\Gamma)$ to second order in the
velocities, as in eq. (\ref{lw}), gives the Darwin Lagrangian $L_D$ describing 
 the interaction of a pair of oppositely charged particles$^{[15],[16]}$
\begin{equation}
L_D = {e^2\over 4\pi R} - {1\over 2} {e^2\over 4\pi R} \left[\vec v_1 \cdot
\vec v_2 + {(\vec v_1 \cdot \vec R) (\vec v_2 \cdot \vec R)\over R^2} \right].
\label{dl}
\end{equation}

In the dual description we consider first the inhomogeneous Maxwell equations,
which we write in the form:
\begin{equation}
-\partial^\beta {\epsilon_{\alpha\beta\sigma\lambda} G^{\sigma \lambda}\over 2}
= j_\alpha ,
\label{im}
\end{equation}
where $G_{\mu\nu}$ is the dual field tensor composed of the electric
displacement vector $\vec D$ and the magnetic field vector $\vec H$:
\begin{equation}
G_{0k} \equiv H_k ,\qquad G_{\ell m} = \epsilon_{\ell mn} D^n.
\label{e310}
\end{equation}
Next we express the charged particle current in eq. (\ref{im}) as the
divergence of
a polarization tensor  $G_{\mu\nu}^S$, the Dirac string tensor, representing a
moving line of polarization running from the negatively charged to the
positively charged particle, namely$^{[14]}$
\begin{equation}
G_{\mu\nu}^S (x) = - e \epsilon_{\mu\nu \alpha\beta} \int d\sigma \int d \tau
{\partial y^\alpha\over\partial\sigma} {\partial y^\beta\over\partial\tau}
\delta (x - y(\sigma,\tau)),
\label{gs}
\end{equation}
where $y^\alpha(\sigma,\tau)$ is a world sheet with boundary $\Gamma$ swept out
by the Dirac string.  Then $^{[1]}$
\begin{equation}
 - \partial^\beta {\epsilon_{\alpha\beta\sigma\lambda}
G^{S\sigma\lambda}(x)
\over 2}=j_\alpha (x) \,
\label{ims}
\end{equation}
and the solution of eq. (3.8) is
\begin{equation}
G_{\mu\nu} = \partial_\mu C_\nu - \partial_\nu C_\mu + G_{\mu\nu}^S,
\label{gmn}
\end{equation}
which defines the magnetic variables (the dual potentials $C_\mu$).  (With eqs.
(3.11) and (\ref{gmn}) the inhomogeneous Maxwell equations become Bianchi
identities.)

The homogeneous Maxwell equations for $\vec E$ and $\vec B$, which we write in
the form
\begin{equation}
\partial^\alpha (\mu G_{\alpha\beta}) = 0,
\label{hm}
\end{equation}
where $\mu = {1\over\epsilon}$ is the magnetic susceptibility, now become
dynamical equations for the dual potentials.  These equations can be obtained
by varying $C_\mu$ in the Lagrangian
\begin{equation}
{\cal L}_C (G_{\mu\nu}^S) = - {1\over 4} \mu G_{\mu\nu} G^{\mu\nu},
\label{lc}
\end{equation}
where $G_{\mu\nu}$ is given by eq. (\ref{gmn}).  This Lagrangian 
provides the
dual (magnetic) description of the Maxwell theory (3.2).
  In the dual description the Wilson loop $W(\Gamma)$
 is  given by 
\begin{equation}
W (\Gamma) \equiv {\int {\cal D}C_\mu e^{i\int dx [{\cal L}_C (G_{\mu\nu}^S) +
{\cal L}_{GF}]}\over \int {\cal D} C_\mu e^{i\int dx [{\cal L}_C
(G_{\mu\nu}^S= 0) + {\cal L}_{GF}]}}.
\label{wc}
\end{equation}

Evaluating the functional integral (\ref{wc}) by completing the square gives
\begin{equation}
W(\Gamma) = e^{-{i\mu\over 4} \int dx G_{\alpha\beta}^S (x) G^{\alpha\beta}
(x)},
\label{wcp}
\end{equation}
where $G^{\alpha\beta} (x)$ is the dual field tensor (\ref{gmn}) with $C_\mu =
C_\mu^D$
determined from the solution of eq. (\ref{hm}), which has the explicit form
\begin{equation}
\partial^\alpha \mu 
 (\partial_\alpha C^D_\beta - \partial_\beta C^D_\alpha) = -
\partial^\alpha \mu G_{\alpha\beta}^{S}.
\label{mc}
\end{equation}
Inserting
\begin{equation}
G^{\alpha\beta} (x) = - {1\over 2} \epsilon^{\alpha\beta\lambda\sigma}
(\partial_\lambda \int dy {\cal D}_{\sigma \beta'} (x-y) j_{\beta'} (y)
- \partial_\sigma \int dy {\cal D}_{\lambda \beta'} (x-y) j_{\beta'}
(y))
\label{gab}
\end{equation}
into (\ref{wcp}), integrating by parts,
and using eq. (3.11), we
obtain the same result (eq. (\ref{wg})
 with ${1\over\epsilon} \rightarrow \mu)$
for
the Wilson loop (3.15) defined in the magnetic description as we had obtained
for the Wilson loop  defined in the electric description.  We then have
two equivalent descriptions at all distances of the electromagnetic interaction
of two charged particles.  (Note, however, that if $\epsilon \rightarrow 0$ at
long distances, then $\mu \rightarrow {1\over\epsilon} \rightarrow \infty$ and
the dual potentials $C_\mu$ determined from eq. (\ref{hm}) are not strongly
coupled at long distances unlike the potentials $A_\mu$ determined from eq.
(\ref{hma}).)

We now return to ${\cal L}_{eff}$, which in absence of quark 
 sources has the form$^{[1]}$
\begin{equation}
{\cal L}_{eff} = 2 tr \left[ - {1\over 4} {\bf G}^{\mu\nu} {\bf G}_{\mu\nu}
+ {1\over 2} ({\cal {\cal D}}_\mu {\bf B}_i)^2\right] - W({\bf B}_i),
\label{leff}
\end{equation}
where
\begin{equation}
{\cal D}_\mu {\bf B}_i = \partial_\mu {\bf B}_i - i g [ {\bf C}_\mu, {\bf
B}_i],
\end{equation}
\begin{equation}
{\bf G}_{\mu\nu} = \partial_\mu {\bf C}_\nu - \partial_\nu {\bf C}_\mu - i g
[{\bf C}_\mu, {\bf C}_\nu],
\label{dft}
\end{equation}
\[
g = {2\pi \over e}, \]
${\bf C}_\mu$ and ${\bf B}_i$ are $SU(3)$ matrices,
and $W ({\bf B}_i)$ is the Higgs potential which has a minimum at non-zero 
values ${\bf B}_{0i}$ which have the color structure
\begin{equation}
{\bf B}_{01} =  B_0 \lambda_7 ,
\quad {\bf B}_{02} = B_0 (-\lambda_5),\quad {\bf
B}_{03} = B_0 \lambda_2.
\end{equation}

The three matrices $\lambda_7, -\lambda_5$ and $\lambda_2$ transform as a $j =
1$ irreducible representation of an $SU(2)$ subgroup of $SU(3)$  and as there
is no $SU(3)$ transformation which leaves all three ${\bf B}_{0i}$
invariant the dual 
$SU(3)$ gauge symmetry is  completely broken and the  eight Goldstone bosons
become the longitudinal components of the now massive ${\bf C}_\mu$.

The basic manifestation of the dual superconducting properties of ${\cal
L}_{eff}$ is that it generates classical equations of motion having
solutions$^{[17]}$
carrying a unit of $Z_3$ flux confined in a narrow tube along the $z$ axis
(corresponding to having quark sources at $z = \pm \infty$).  (These solutions
are dual to Abrikosov--Nielsen--Olesen magnetic vortex solutions$^{[18]}$ in a
superconductor).  We briefly describe these classical solutions here in order
to specify the  color structures that enter into the subsequent
treatment of the dual theory with quark sources which is not restricted to the
classical approximation.
We choose a gauge where the dual potential
 is proportional to the
hypercharge matrix $Y = {\lambda_8\over\sqrt{3}}$:
\begin{equation}
{\bf C}_\mu = C_\mu Y.
\end{equation}
As a consequence the non-Abelian terms in the expression  (\ref{dft}) for
the
dual field tensor ${\bf G}_{\mu\nu}$ vanish.

We choose Higgs Fields ${\bf B}_i$ having the following color structure:
\begin{eqnarray}
{\bf B}_1 &=& B_1 (x) \lambda_7 + \bar B_1 (x) (-\lambda_6)\nonumber\\
{\bf B}_2 &=& B_2 (x) (-\lambda_5) + \bar B_2 (x) \lambda_4 \nonumber\\
{\bf B}_3 &=& B_3 (x) \lambda_2 + \bar B_3 (x) (-\lambda_1).
\end{eqnarray}
With this ansatz the Higgs potential $W$ turns out to be
\begin{eqnarray}
W & = &{2\over 3} \lambda \left\{  11 \big [
(|\phi_1|^2 - B_0^2)^2 +
 (|\phi_2|^2 - B_0^2)^2
+ (|\phi_3|^2 - B_0^2)^2 \big ]  + {7}
(|\phi_1|^2 + |\phi_2|^2 + |\phi_3|^2 - 3
B_0^2)^2\right\}\qquad,\nonumber\\
& & \quad
\label{wb}
\end{eqnarray}
where
\begin{equation}
\phi_i (x) = B_i (x) - i \bar B_i (x) .
\end{equation}
Using  (3.23) and (3.24) we also find
\begin{equation}
2 tr \sum_{i}
({\cal D}_\mu {\bf B}_i)^2 = 4 |(\partial_\mu - i g C_\mu) \phi_1|^2 +
4|(\partial_\mu - ig C_\mu) \phi_2|^2 + 4|\partial_\mu \phi_3|^2.
\end{equation}
Since $\phi_1$ and $\phi_2$ couple to $C_\mu$ in the same way and $\phi_3$ does
not couple to $C_\mu$ at all, we can choose $\phi_1 = \phi_2 = \phi \equiv B -
i
\bar B$, and $\phi_3 = B_3$, so that
\begin{equation}
2 tr \sum_{i}
({\cal D}_\mu {\bf B}_i)^2 = 8|(\partial_\mu - ig C_\mu) \phi|^2 + 4
(\partial_\mu
B_3)^2.
\label{tdb}
\end{equation}

At large distances from the center of the flux tube, using cylindrical
coordinates $\rho,\theta, z$ we have the boundary conditions:
\begin{equation}
\vec C \rightarrow - {\hat e_\theta\over g\rho}, ~\phi \rightarrow B_0
e^{i\theta}, ~B_3 \rightarrow B_0, ~{\rm as}~ \rho \rightarrow \infty.
\end{equation}
The non-vanishing of $B_{0}$ produces a color monopole current confining the
electric color flux.  The line integral of the dual potential around a large
loop surrounding the $z$ axis measures the electric color flux, just as the
corresponding line integral of the ordinary vector potential measures the
magnetic flux in a superconducting vortex.  Since the dual potential is along a
single direction in color space 
  path ordering is unnecessary
and the boundary condition (3.29) for $\vec {\bf C}$ gives
\begin{equation}
e^{- ig \oint_{loop} \vec {\bf C} \cdot  d \vec \ell} = e^{2\pi i Y} =
e^{2\pi\left({i\over 3}\right)},
\end{equation}
which manifests the unit of $Z_3$ flux in the tube.  (A continuous deformation
in $SU(3)$ of our particular solution into a non-Abelian configuration will
leave unchanged the path ordered integral $P\exp (-ig \int \vec {\bf C}
 \cdot d\vec
\ell) = e^{2\pi{i\over 3}}$).  The energy per unit length in this flux tube is
the string tension $\sigma$.  The quantity $g^2/\lambda$ plays the role of a
Landau--Ginzburg parameter. Its value  can be obtained by relating
 the difference between the energy density at a large distance 
from the flux tube and the energy density at its center to the gluon 
 condensate. This procedure gives $g^2/\lambda=5$ (which is near
the border between type I and type II superconductors). We get from the
numerical integration of the static field equations$^{[17]}$
\begin{equation}
\sigma \approx 1.1 (24 B_0^2).
\end{equation}
We
are left with two free parameters in ${\cal L}_{eff}$, which we take to be
$\alpha_s = {e^2\over 4\pi} = {\pi\over g^2}$ and the string tension
$\sigma$.

To couple ${\bf C}_\mu$
 to a $q\bar q$ pair separated by a finite distance we must
represent quark sources by a Dirac string tensor ${\bf G}_{\mu\nu}^S$.  We
choose  the dual potential to have the same color structure (3.23) as the flux
tube solution.  Then ${\bf G}_{\mu\nu}^S$ must also be proportional to the
hypercharge matrix
\begin{equation}
{\bf G}_{\mu\nu}^S = Y G_{\mu\nu}^S,
\end{equation}
where $G_{\mu\nu}^S$ is given by eq. (\ref{gs}), so that one unit of $Z_3$ flux
flows along the Dirac string connecting the quark and anti--quark.   We then
couple quarks 
 by replacing ${\bf G}_{\mu\nu}$ in ${\cal L}_{eff}$ 
(\ref{leff})
by
\begin{equation}
{\bf G}_{\mu\nu} = Y G_{\mu\nu},
\end{equation}
where
\begin{equation}
G_{\mu\nu} = \partial_\mu C_\nu - \partial_\nu C_\mu + G_{\mu\nu}^S.
\label{gmn2}
\end{equation}

Inserting (3.33) into (\ref{leff}) and using eq. (\ref{tdb}) then yields the
Lagrangian
${\cal L}_{eff} (G_{\mu\nu}^S)$ coupling dual potentials to classical quark
sources moving along trajectories $\vec z_1 (t)$ and $\vec z_2 (t)$:
\begin{equation}
{\cal L}_{eff} (G_{\mu\nu}^S) = - {4\over 3} {(\partial_\mu C_\nu -
\partial_\nu
C_\mu + G_{\mu\nu}^S)^2\over 4} + {8 |(\partial_\mu - ig C_\mu) \phi|^2\over 2}
+ {4(\partial_\mu B_3)^2\over 2} - W,
\label{leff2}
\end{equation}
where $W$ is given by  (\ref{wb}) with $\phi_1 =
\phi_2 = \phi, \phi_3 = B_3$.

It is useful, as in (3.9), 
to decompose $G_{\mu\nu}$ into its color electric components $\vec
D$ and color magnetic  components $\vec H$.  Similarly we
decompose $G_{\mu\nu}^S$ into its polarization components $\vec D_S$ and its
magnetization components $\vec H_S$:
\begin{equation}
 D_S^k = {1\over 2} \epsilon_{kmn} G^{Smn},  ~H_S^k = G_{0k}^S.
\end{equation}
Then eq. (3.34) becomes
\begin{equation}
\vec D = - \vec\nabla \times \vec C + \vec D_S, \quad \vec H = - \vec \nabla
C_0 - {\partial \vec C\over \partial t} + \vec H_S.
\end{equation}
The Lagrangian density ${\cal L}_{eff}(G_{\mu\nu}^S)$ (3.35)
 can then be written as the
sum of an ``electric'' part ${\cal L}_0$ and a ``magnetic'' part ${\cal L}_2$.
i.e.,
\begin{equation}
{\cal L}_{eff}(G_{\mu\nu}^S) = {\cal L}_0 + {\cal L}_2,
\end{equation}
where
\begin{equation}
{\cal L}_0 = - \left\{{2\over 3} \vec D^2 + 4|(\vec\nabla + ig \vec C) \phi|^2
+ 2 (\vec\nabla B_3)^2 + W\right\},
\end{equation}
and
\begin{equation}
{\cal L}_2 = {2\over 3} \vec H^2 + 4| (\partial_0 - ig C_0) \phi|^2 +
2(\partial_0 B_3)^2 ,
\end{equation}
and all terms involving time derivatives appear only in ${\cal L}_2$.

We denote by $W_{eff} (\Gamma)$ the Wilson loop of the dual theory,
i.e.,
\begin{equation}
  W_{eff} (\Gamma) =
   {
   \int {\cal D} C_\mu {\cal D} \phi {\cal D} B_3
    e ^ {i \int dx [ {\cal L}_{eff} (G_{\mu\nu}^S) + {\cal L}_{GF} ] }
  \over
  \int {\cal D} C_\mu {\cal D} \phi {\cal D} B_3
    e ^ {i \int dx [ {\cal L}_{eff} (G_{\mu\nu}^S=0) + {\cal L}_{GF} ] }
   }.
  \label{weff}
\end{equation}
The functional integral $W_{eff} (\Gamma)$ eq. (3.41) determines in the dual
theory the same physical quantity as $W(\Gamma)$ in Yang--Mills theory, namely
the action for a quark-antiquark pair moving along classical trajectories.  The
coupling in ${\cal L}_{eff} (G_{\mu\nu}^S)$ of dual potentials to Dirac strings
plays
the role in the expression (\ref{weff}) for $W_{eff} (\Gamma)$ of the explicit
Wilson
loop  integral
 $e^{-ie\oint_\Gamma dx^\mu A_\mu (x)}$ in the expression (2.2) for
$W(\Gamma)$.\footnote{We emphasize the distinction between $W_{eff} (\Gamma)$
and the Wilson loop  of the dual theory defined as an average of
$e^{ig \oint \vec C \cdot d\vec \ell}$.  This dual Wilson loop would describe
the
interaction of a monopole antimonopole pair.  For large loops the dual Wilson
loop satisfies a perimeter law in accordances with 't Hooft's
observation.}

The assumption that the dual theory describes the long distance $q \bar q$
interaction in Yang--Mills theory then takes the form:
\begin{equation}
W(\Gamma) = W_{eff} (\Gamma) , ~{\rm for ~large ~loops ~\Gamma}.
\label{ass}
\end{equation}
Large loops means that the size $R$ of the loop is
large compared to the inverse mass of the Higgs particle (monopole field)
$\phi$.  Furthermore since the dual theory is
weakly coupled at large distances we can evaluate $W_{eff} (\Gamma)$ via a semi
classical expansion to which the classical configuration of dual potentials and
monopoles gives the leading contribution.  This then allows us to picture heavy
quarks (or constituent quarks) as sources of a long distance classical field of
dual gluons determining the heavy quark potential.  Thus, in a certain sense
the dual gluon fields  $G_{\mu\nu}$ mediate the heavy quark interaction just as
the electromagnetic field mediates the electron positron interaction.

Using the duality hypothesis, we replace $W(\Gamma)$ by $W_{eff}
(\Gamma)$ in eqs. (\ref{lw})-(2.6) to obtain expressions for $V_0 (R)$
and $V_{VD}$
in the dual theory as the zero
order and quadratic terms in the expansion of $i\log W_{eff} (\Gamma)$ for
small velocities ${\dot{\vec z}}_1$ and ${\dot{\vec z}}_2$, i.e., the
interaction Lagrangian $L_I$, calculated in the dual theory, is obtained from
the equation
\begin{equation}
i \log W_{eff} (\Gamma) = - \int_{t_{i}}^{t_{t}} dt L_I (\vec z_1, \vec
z_2, {\dot{\vec z}}_1, {\dot{\vec z}}_2).
\label{lweff}
\end{equation}

{\bf Remark}

There has been a recent revival of interest in the role of electric magnetic
duality due to the work of Seiberg,$^{[7]}$ Seiberg and Witten$^{[8]}$ and
others on super symmetric non-Abelian gauge
theories.  Seiberg$^{[7]}$ considered $SU(N_c)$
gauge theory with $N_f$ flavors of massless quarks.  Although he did not
exhibit an explicit duality transformation he inferred  the complete
structure of the magnetic gauge group and hence the associated massless
particle content of the dual Lagrangian.  For a certain range of $N_f$ the dual
theory is weakly coupled at large distances and hence the low energy spectrum
of the theory consists just of the massless particles of the dual Lagrangian.
Since this dual ``magnetic'' Lagrangian describes the same low energy physics
as the original Lagrangian, the particle spectrum, mirroring the magnetic gauge
group, must appear as composites of the original ``electric'' gauge degrees of
freedom.   For $N_f = N_c + 1$ the dual gauge group is completely broken, 
the associated dual gauge bosons become massive and the quarks of
the original theory are confined.

There are obvious differences between Seiberg's example, where the number of
massless fermions plays an essential role, and the example of Yang--Mills
theory where neither the original theory nor the proposed dual Lagrangian
${\cal L}_{eff}$ contains fermions.  Here confinement manifests itself via the
development of a linear potential between heavy quark sources, whereas in the
supersymmetric models confinement manifests itself via the realization of the
hadron spectrum as composites of the original quark variables.  In the
supersymmetric model these hadrons are massless and as usual the production  of
these particles prevents the development of a linear potential.  However, all
the gauge bosons of the dual theory are massive and the coupling of the pure
gauge sector to quark sources would produce a long distance linear potential
between these sources.  The common feature of Seiberg's supersymmetric model,
where duality is "inferred",
 and Yang--Mills theory, where duality is assumed,
is that in both cases the dual gluons receive mass via a Higgs mechanism which
is the essential element of the dual superconductor mechanism.

\section{The Potential $V_{q\bar q}$ in the Dual Theory}

We now express the spin dependent heavy quark potential $V_{SD}$ (2.8) in
terms of quantities of the dual theory.  As a first step we find relations
of matrix elements of the dual field tensor $G_{\mu\nu}$
to variations of $W_{eff} (\Gamma)$ which are analogous to eq. (2.20) relating
$\langle\!\langle F_{\mu\nu}\rangle\!\rangle$ to variations in $W(\Gamma)$.
Consider the variation in $W_{eff} (\Gamma)$ produced by the change
\begin{equation}
G_{\mu\nu}^S (x) \rightarrow G_{\mu\nu}^S (x) + \delta G_{\mu\nu}^S (x).
\end{equation}
{}From eq. (\ref{weff}) we find that the corresponding variation $\delta
W_{eff}
(\Gamma)$ is given by
\begin{equation}
\delta i\log W_{eff}(\Gamma) = {4\over 3}  \int dx {\delta G_{\mu\nu}^S (x)\over
2}
\langle\!\langle G^{\mu\nu} (x) \rangle\!\rangle_{eff},
\label{dweff}
\end{equation}
where
\begin{equation}
\langle\!\langle f(C_\mu, \phi, B_3)\rangle\!\rangle_{eff} \equiv {\int {\cal
D}
C_\mu {\cal D} \phi {\cal D} B_3 e^{i\int dx
({\cal L}_{eff} (G_{\mu\nu}^S) + {\cal L}_{GF})} f (C_\mu, \phi, B_3)\over \int
{\cal D} C_\mu {\cal D} \phi {\cal D} B_3 e^{i\int dx
 ({\cal L}_{eff}
(G_{\mu\nu}^S) + {\cal L}_{GF})}}.
\end{equation}

Using  (\ref{gs}) to express the variation of $G_{\mu\nu}^S$ in terms of the
variation of the world sheet $y^\mu (\sigma,\tau)$, we obtain
\[
\int dx  {\delta G_{\mu\nu}^S\over 2}
 (x) \langle\!\langle G^{\mu\nu} (x)
\rangle\!\rangle_{eff} =\]
\begin{equation}
- {e\over 2} \epsilon_{\mu\nu \lambda\alpha} \int_{t_{i}}^{t_{f}} d\tau \left[
\delta z_1^\alpha {\partial z_1^\lambda\over \partial\tau} \langle\!\langle
G^{\mu\nu} (z_1)
\rangle\!\rangle_{eff} - \delta z_2^\alpha {\partial z_2^\lambda\over
\partial\tau} \langle\!\langle
G^{\mu\nu} (z_2)\rangle\!\rangle_{eff}\right].
\label{gsg}
\end{equation}
The right hand-side of eq. (\ref{gsg}) arises from varying the boundary of the
Dirac
sheet.  The variation of the interior of the sheet produces a contribution
proportional to the monopole current $j_\nu^{MON}$:
\begin{equation}
 j_\nu^{MON}(x) \equiv 
\partial^\mu G_{\mu\nu} (x).
\end{equation}
(See eq. A.52 of Reference 1 for details).  This gives no additional
contribution to eq. (\ref{gsg}) since the monopole current
must vanish on the Dirac sheet, so that no monopole can pass through the Dirac
string connecting the charged particles.  This latter assertion is just the
dual of Dirac's condition for the consistency of a theory containing both
electric charges and monopoles$^{[14]}$.

Defining $dz^\lambda \equiv d\tau {\partial z^\lambda\over\partial\tau}$, we
can
then write eq. (\ref{gsg}) as
\[
\int dx {\delta G_{\mu\nu}^S (x)
\over 2} \langle\!\langle G^{\mu\nu} {(x)}
\rangle\!\rangle _{eff} =\]
\begin{equation}
- {e}\int (\delta z_1^\alpha d z^\lambda_1 \langle\!\langle \hat
G_{\lambda\alpha}
(z_1)\rangle\!\rangle_{eff} -\delta z_2^\alpha dz_2^\lambda \langle\!\langle
\hat G_{\lambda\alpha}
(z_2)\rangle\!\rangle_{eff}),
\end{equation}
where
\begin{equation}
\hat G_{\mu\nu} (x) \equiv {1\over 2} \epsilon_{\mu\nu\lambda\sigma}
G^{\lambda\sigma} (x).
\end{equation}
Choosing a variation which vanishes on the curve $\Gamma_2$, we
obtain
\begin{equation}
\delta i \log W_{eff} (\Gamma) = - {4\over 3} e \int {\delta S^{\mu\nu}
(z_1) \over 2}
 \langle\!\langle \hat G_{\mu\nu} (z_1) \rangle\!\rangle_{eff},
\label{delweff}
\end{equation}
where $\delta S^{\mu\nu} (z_1)$ is given by eq. (\ref{smn}).  Eqs. (4.2) and
(4.8) can be written as
\begin{equation}
{\delta i \log W_{eff} (\Gamma)\over \delta S^{\mu\nu} (z_1)} 
 = - {4\over 3} {e} \langle\langle \hat
G_{\mu\nu} (z_1) \rangle\rangle_{eff}=
 - {e\over 2}\,
 \varepsilon_{\mu \nu \lambda \sigma} {\delta i \log W_{eff}
(\Gamma)\over \delta G_{\lambda \sigma}^S(z_1) },
\end{equation}
which is the dual theory analogue of eq. (\ref{e20}).  The duality assumption
(\ref{ass}) then gives a corresponding relation between matrix elements:
\begin{equation}
\langle\!\langle F_{\mu\nu} (z_1)\rangle\!\rangle = 
 {4\over 3} \langle\!\langle
\hat G_{\mu\nu} (z_1)\rangle\!\rangle_{eff}.
\label{fmn}
\end{equation}
Eq. (4.10) gives a correspondence between local quantities in
Yang--Mills theory and in the dual theory.  The utility of electric-magnetic
duality is that for large loops semi-classical configurations dominate the
right hand side of
eq. (\ref{fmn}) in contrast to the rapidly fluctuating configurations of
Yang--Mills
potential which contribute to the left hand side.   Eq. (\ref{fmn}) breaks up
into
its electric and magnetic components:
\begin{equation}
-\langle\!\langle F_{mn} \rangle\!\rangle = {4\over 3} \epsilon_{mn\ell}
\langle\!\langle H^\ell \rangle\!\rangle_{eff},
\label{heff}
\end{equation}
\begin{equation}
\langle\!\langle F_{0\ell} \rangle\!\rangle =  {4\over 3} \langle\!\langle
D_\ell \rangle\!\rangle_{eff}.
\label{deff}
\end{equation}
or equivalently,
\begin{eqnarray}
\langle\!\langle \hat F_{0\ell} \rangle\!\rangle &=& {4\over 3}
\langle\!\langle H_\ell \rangle\!\rangle_{eff},\\
\langle\!\langle \hat F_{mn} \rangle\!\rangle &=& {4\over 3} \epsilon_{mnk}
\langle\!\langle D_k\rangle\!\rangle_{eff}.
\end{eqnarray}

Using eqs. (4.13) and (4.14) in eq. (\ref{vls}) gives the following expression
for
$V_{LS}^{MAG}$ in the dual theory:
\begin{equation}
V_{LS}^{MAG} = - \sum_{j=1}^2 {4\over 3} {e_j\over m_j} \vec S_j \cdot
(\langle\!\langle \vec H(z_j) \rangle\!\rangle_{eff} - \vec v_j \times
\langle\!\langle \vec D(z_j) \rangle\!\rangle_{eff}),
\label{nvls}
\end{equation}
where $e_1 = e$ and $e_2 = - e$.  Note that $\langle\!\langle \vec H
\rangle\!\rangle_{eff} - \vec v_j \times \langle\!\langle \vec D
\rangle\!\rangle_{eff}$ is the magnetic field at the $j$th quark in the
comoving Lorentz frame, $V_{LS}^{MAG}$ the magnetic interaction of this
field with a quark having a $g$ factor 2.    The fact that heavy quarks
interact with a Dirac magnetic moment is
a consequence of the ${1\over m}$ expansion $^{[2]}$
 for the $q\bar q$ Green's
function upon which this analysis is based.

To evaluate $V_{Thomas}$ (2.13) we note
from  (\ref{deff}) that
\begin{eqnarray}
& & {e\over 2m_1^2} \int dz_1^\mu S_1^\ell \epsilon^{\ell k r} p_1^k
\langle\!\langle F_{\mu r}(z_1) \rangle\!\rangle\nonumber \\
&=& {4\over 3} {e\over 2m_1^2} \int dt \vec S_1 \cdot \vec p_1 \times
 \langle\!\langle \vec D (z_1) \rangle\!\rangle_{eff} ,
\label{e415}
\end{eqnarray}
and  obtain
\begin{equation}
V_{Thomas} = - {1\over 2} \sum_{j = 1}^2 {4\over 3} {e_j\over m_j} S_j \cdot
(\vec v_j \times \langle\!\langle \vec D (z_j)\rangle\!\rangle_{eff}).
\label{nvt}
\end{equation}
 The expression (4.17) is  the
contribution to the potential due to the precession of the axis of the comoving
frame.  In appendix A it is shown that
(\ref{nvt}) can be written in the usual form
\begin{equation}
V_{Thomas} = {1\over 2m_1} {1\over R} {\partial V_0\over \partial R} \vec S_1
\cdot \vec v_1 \times \vec R - {1\over 2m_2} {\partial V_0\over\partial R} \vec
S_2 \cdot \vec v_2 \times \vec R.
\label{evt}
\end{equation}
Eq. (\ref{evt}) is essentially a kinematic relation and is independent of the
dynamics
of Yang--Mills theory.  On the other hand $V_{LS}^{MAG}$, eq. (\ref{nvls}),
depends
upon the dynamics and cannot be expressed solely in terms of the central
potential.

To express $V_{SS}$ (2.17) in terms of quantities involving the dual
theory we need the following:
\[ i e^2 \{\langle\!\langle \hat F_{\ell 0} (z_j) \hat F_{k0}
(z_{j^\prime}^\prime)\rangle\!\rangle - \langle\!\langle \hat F_{\ell 0}
(z_j) \rangle\!\rangle \langle\!\langle \hat F_{k0} (z_{j^\prime}^\prime)
\rangle\!\rangle\}\]
\begin{equation}
= {4\over 3} e^2 {\delta \langle\!\langle  H_k
(z_j)\rangle\!\rangle_{eff}\over \delta  H_{S\ell}
(z_{j^\prime}^\prime)}.
\label{hhs}
\end{equation}
To obtain (\ref{hhs}) we use eqs. (\ref{e21}) and (4.13) and the equation
\begin{equation}
{\delta \langle\!\langle H_k(z_j)
 \rangle\!\rangle_{eff}\over \delta G_{0\ell}^S(z_{j^\prime})} =
- {\epsilon_{\ell mn}\over 2} {\delta \langle\!\langle H_k(z_j)
 \rangle\!\rangle_{eff}\over
\delta S^{mn}(z_{j^\prime})}.
\label{hs}
\end{equation}
(Compare eq. (4.9)).  Using eq. (4.19) in   (2.17), we
obtain
\begin{equation}
\int dt V_{SS} = - {1\over 2} \sum_{j,j'=1}^2 {e^2\over m_j m_{j'}} T_s
\int_{\Gamma_{j}} dt \int_{\Gamma_{j'}} dt' S_j^\ell S_{j'}^k \left({4\over 3}
{\delta \langle\!\langle H_k (z_j) \rangle\!\rangle_{eff}\over \delta H_{S\ell}
(z_{j^\prime}^\prime)}\right).
\label{nvss}
\end{equation}
The factor multiplying $S_j^\ell S_{j'}^k$ is symmetric in $k$ and $\ell$ and
hence  the terms in eq. (\ref{nvss}) where $j = j'$ involve the
combination
\[
{S_j^\ell S_j^k + S_j^k S_j^\ell\over 2} = {1\over 4} \delta_{k\ell}.\]
Eq. (\ref{nvss}) then becomes
\[
\int dt V_{SS} = - {4\over 3} \sum_{j=1}^2 {e^2\over 8m_j^2} \int_{\Gamma_{j}}
dt \int_{\Gamma_{j}} dt' {\delta \langle\!\langle H_k
(z_j)\rangle\!\rangle_{eff}\over \delta H_{Sk}
(z_{j^\prime}^\prime)}\]
\begin{equation}
- {4\over 3} \left( {e^2\over m_1 m_2}\right) \int_{\Gamma_{1}} dt
\int_{\Gamma_{2}} dt' S_1^k S_2^\ell {\delta \langle\!\langle H_k (z_1)
\rangle\!\rangle_{eff}\over \delta H_{S\ell} (z_2^\prime)}.
\label{nnvss}
\end{equation}
The first term in eq. (\ref{nnvss}) is a spin independent velocity independent
contribution to the potential proportional to inverse square of the quark
masses while the second term in eq. (\ref{nnvss}) yields a spin-spin
interaction of
the expected structure.

Finally, let us come to $V_{Darwin}$ (2.15) and note that
\begin{equation}
\langle\!\langle D^\nu F_{\nu\mu}(z_j) \rangle\!\rangle = \partial^\nu
\langle\!\langle F_{\nu\mu} {(z_j)} \rangle\!\rangle.
\label{dfmn}
\end{equation}
The derivative of the Wilson loop occuring in the definition (2.10) of
$\langle\!\langle F_{\nu\mu} (x)\rangle\!\rangle$ yields the Yang--Mills
potential $A_\nu$ appearing in $D^\nu F_{\nu\mu}$. Using (4.10) we obtain 
\footnote{Notice that
$\langle\!\langle F_{\mu\nu}(z) \rangle\!\rangle$ depends
 not only on the point $z$ but on the entire Wilson loop.  So in order for eq.
 (4.23) to make sense one has to use the appropriate definition of derivative.
Given a functional $\Phi_{[\gamma_{ab}]}$ of the curve
 $\gamma_{ab}$ with ends $a$ and $b$, under general regularity condition the
 variation of $\Phi$ consequent to an infinitesimal modification of the curve
 $\gamma \rightarrow \gamma + \delta \gamma$ can be expressed as the sum of
 various terms proportional respectively  to $\delta a, \delta b$ and to the
  elements $\delta S_{\rho\sigma} (x)$ of the surface swept by the
 curve.  Then the derivatives $\partial/\partial a^\rho, \partial/\partial
 b^\rho$ and $\delta/\delta S^{\rho\sigma} (x)$ are defined by the equation
 $\delta\Phi = \partial\Phi/\partial a^\rho \delta a^\rho + \partial
\Phi/\partial
 b^\rho \delta b^\rho + \int_\gamma \delta S^{\rho\sigma} (x) \delta
 \Phi/\delta S^{\rho\sigma} (x)$.  In our case this would amount to put
 naively $\partial/\partial z^\rho P f (\int_z^b dx^\mu A_\mu (x)) = - P f'
 (\int_z^b dx^\mu A_\mu (x)) A_\rho (z)$ and $\partial/\partial z^\rho
 \int_a^z dx^\mu A_\mu (x) = A_\rho (z) Pf' (\int_a^z dx^\mu A_\mu
 (x))$.}
\begin{equation}
\int V_{Darwin} dt = - {4\over 3} \sum_j {e\over 8m_j^2} \int_{\Gamma_{j}}
dx^\mu \partial^\nu \langle\!\langle \hat G_{\nu\mu}(x)
 \rangle\!\rangle_{eff}.
\label{nvd}
\end{equation}
For an alternative expression for $V_{Darwin}$ based on (eq. A.7) of Ref. [4]
see eq. B.3 of Appendix B.

\section{The Classical Approximation for $V_0 (R)$ and $V_{VD}$}

In the classical approximation eq. (3.41) becomes
\begin{equation}
i \log W_{eff} = - \int dx {\cal L}_{eff} (G_{\mu\nu}^S),
\end{equation}
where ${\cal L}_{eff} (G_{\mu\nu}^S)$  (\ref{leff2}) is evaluated
at the solution of the classical equations of motion:
\begin{equation}
\partial^\alpha (\partial_\alpha C_\beta - \partial_\beta C_\alpha) = -
\partial^\alpha G_{\alpha \beta}^S + j_\beta^{MON},
\label{e52}
\end{equation}
\begin{equation}
(\partial_\mu - i g C_\mu)^2 \phi = - {1\over 4} {\delta W\over\delta\phi^*} ,
\end{equation}
and
\begin{equation}
\partial^2 B_3 = - {1\over 4} {\delta W\over\delta B_3} ,
\end{equation}
where the monopole current $j_\mu^{MON}$ is
\begin{equation}
j_\mu^{MON} = - 3i g [\phi^* (\partial_\mu - ig C_\mu) \phi - \phi
(\partial_\mu + i g C_\mu) \phi^*].
\label{jm}
\end{equation}
As a result of the classical approximation all quantities in brackets
 are replaced by their classical values
\begin{equation}
\langle\!\langle G_{\mu\nu}(x) \rangle\!\rangle_{eff} = G_{\mu\nu} (x).
\label{geff}
\end{equation}
  The electric and
magnetic components of eq. (5.6) are
\begin{equation}
\langle\!\langle \vec D (x) \rangle\!\rangle_{eff} = \vec D(\vec x), \quad
\langle\!\langle \vec H (x) \rangle\!\rangle_{eff} = \vec H(\vec x),
\end{equation}
where $\vec D$ and $\vec H$ are the color electric and magnetic fields
respectively given in terms of the dual potentials by eq. (3.37).

We choose the Dirac string to be a straight line $L$ connecting the quarks.  As
$\vec x$ approaches  the string,
 $\phi (x) \rightarrow 0, C_\mu (x) \rightarrow
C_\mu^D (x)$, satisfying eq. (\ref{mc}).  As $\vec x \rightarrow \infty, \phi
(x)
\rightarrow B_0, C_\mu (x) \rightarrow 0$, in contrast with the large distance
boundary condition for the infinite flux tube.  We can then choose $\phi (x)$
to be real so that
\begin{equation}
\phi (x) = B(x), \quad j_\mu^{MON} (x) = - 6 g^2 C_\mu B^2.
\end{equation}

Consider first the case of static quarks, $\vec v_1 = \vec v_2 = 0$.  Then the
scalar potential $C_0$ and the color magnetic field $\vec H$ vanish, and ${\cal
L}_{eff}$ reduces to ${\cal L}_0$ eq. (3.39)  which yields the static
potential:
\begin{equation}
V_0 (R) = - \int d\vec x {\cal L}_0,
\label{v0}
\end{equation}
where ${\cal L}_0$ is evaluated at the static solution of eqs. (5.2) - (5.4),
which have the following form in this case:
\begin{equation}
- \vec\nabla \times (\vec\nabla \times \vec C) - 6g^2 B^2 \vec C = - \vec
\nabla \times \vec D_S,
\label{dc}
\end{equation}
\begin{equation}
 (-\nabla^2 + g^2 \vec C^2) B = - {2\lambda\over 3} B (25B^2 + 7B_3^2 - 32
B_0^2),
\label{db}
\end{equation}
and
\begin{equation}
- \nabla^2 B_3 = - {4\lambda\over 3} B_3 (7B^2 + 9B_3^2 - 16B_0^2),
\label{db3}
\end{equation}
where we have used the explicit form, eq. (\ref{wb}), of W.

To solve eq. (\ref{dc}) it is convenient to write
\begin{equation}
\vec C = \vec C^{ D} + \vec c,
\end{equation}
where $\vec C^D$ is the Dirac potential 
 satisfying the static form of eq. (\ref{mc}) namely:
\begin{equation}
- \vec \nabla \times (\vec\nabla\times\vec C^D) = - \vec\nabla\times\vec D_S,
\label{dcd}
\end{equation}
with $\vec D_S$ given by eq. (3.36).  In cylindrical coordinates
with the $z$ axis along the line joining the two quarks at $z = \pm R/2$, eq.
(3.10) gives
\begin{equation}
\vec D_S = e \hat e_z \{ \theta (z - R/2) - \theta (z + R/2)\} \delta (x)
\delta (y),
\end{equation}
which describes the polarization vector for a line of dipoles.  The solution of
eq. (\ref{dcd}) is
\begin{equation}
\vec C^D = \hat e_\phi C^{D} ,
\end{equation}
where
\begin{equation}
C^D = {e\over 4\pi\rho} \left\{{z - R/2\over\sqrt{\rho^2 + (z - R/2)^2}} - {(z
+ R/2)\over \sqrt{\rho^2 + (z+R/2)^2}}\right\}.
\end{equation}
Then
\begin{equation}
\vec c = \hat e_\phi c \, ,
\end{equation}
and eq. (\ref{dc}) becomes the following equation for $c$:
\begin{equation}
(\tilde{\nabla}^2 - 6g^2 B^2) c = 6 g^2 B^2 C^D,
\label{ndc}
\end{equation}
where
\begin{equation}
\tilde{\nabla}^2 f (\rho,z) \equiv {\partial\over\partial\rho}
\left({1\over\rho}
{\partial\over\partial\rho} (\rho f)\right) + {\partial^2 f\over\partial z^2}.
\end{equation}
Eqs. (\ref{db}), (\ref{db3}) and (\ref{ndc}) are three nonlinear equations for
the static
configuration $c, B,$ and $B_3$ with boundary conditions: $c
\rightarrow - C^D, B\rightarrow B_0,~ B_3 \rightarrow B_0$ at large distances;
$c \rightarrow 0, B \rightarrow 0$ for $\vec x$ on $L$.  
These 
equations have been solved in ref.$^{[19]}$
 with the following results:

The monopole current in eq. (\ref{dc}) screens the color electric
field produced by the quark sources  so that as the quark
anti-quark separation
increases the lines of $\vec D$ are compressed from their Coulomb like behavior
at small $R$ to form a flux tube, and thus 
$V_0(R) \rightarrow \sigma R$ at large $R$.
Both
this small $R$ and this large $R$ behavior of the potential have their common
origin in the evolving distribution of the flux of $\vec D$ whose divergence is
fixed by the color electric charge of the quarks $(\vec \nabla \cdot \vec D 
 = \vec
\nabla \cdot \vec D_S$) and whose curl is
determined by the monopole current.  Thus, the dual theory already in the
classical approximation gives a potential which evolves smoothly from the large
$R$ confinement region to the short distance perturbative domain.  This shows
how the dual theory realizes the Mandelstam 't Hooft mechanism.  It does not
describe QCD at shorter distances where radiative corrections giving rise to
asymptotic freedom and a running coupling constant are important.

To calculate the terms in $i \log W_{eff}$ which are quadratic in the quark
velocities we solve the field equations for moving quarks.  To first order in
the velocities the static field distributions follow the quark motion
adiabatically.  The time dependence of $\vec C,B$ and $B_3$ then results from
the
explicit time dependence of $R$.  Furthermore, since $\int d\vec x {\cal L}_0$
generates the static field equations, it is stationary about the  solution to
these equations and remains unchanged to second order in the velocities.
The velocity dependence in the potential then comes from the ``magnetic''
contribution ${\cal L}_2$ which depends quadratically upon
$\partial_0 \vec C, \partial_0 B_3$, and $C_0$, all of which are first order in
the velocities.   The scalar potential $C_0$ satisfies the equation, obtained
from the time component of eq. (\ref{e52}),
\begin{equation}
\nabla^2 C_0 - 6g^2 B^2 C_0 = \vec \nabla \cdot \vec H_S,
\label{dc0}
\end{equation}
valid to first order in the velocities.  With the Higgs field $B(\vec x)$
already determined by the static equations,  eq. (\ref{dc0}) is a linear
equation
for the scalar potential, giving $C_0$ to first order in the velocity.
The velocity dependent potential $V_{VD}$ is then given by
\begin{equation}
V_{VD} = - \int d\vec x {\cal L}_2,
\label{nvvd}
\end{equation}
representing the magnetic color energy due to the fields following the moving
quarks.

 For small $R$ the potential $V_{VD}$ approaches the
velocity
dependent part of the Darwin potential (3.7) (multiplied by the color factor
4/3) because for small $R$ the color magnetic field $\vec H (\vec x)$ becomes
the ordinary Biot--Savart magnetic field.  As $R$
increases the color magnetic field lines are compressed 
 so that  for large separation $V_{VD}$
becomes linear in $R$.  As an example consider the case
in which  two equal mass quarks move in a circular orbit of frequency $\omega$.
Then $\vec v_1 = - \vec v_2 = {\vec\omega \times \vec R\over 2}$, and $V_{VD}$
reduces to
\begin{equation}
V_{VD} = - {1\over 2} I (R) \omega^2,
\end{equation}
which defines the momentum of inertia $I(R)$ of the rotating flux tube
distribution.  Eq. (\ref{nvvd}) evaluated for this configuration of moving
quarks
then determines $I(R)$.  For Large $R$ we find$^{[1]}$
\begin{equation}
\lim_{R\rightarrow\infty} I(R) = {1\over 2} (AR)R^2,
\end{equation}
where
\begin{equation}
A\simeq .21\sigma,
\end{equation}
 determined numerically from eq. (\ref{nvvd}).
By
comparison we note that the moment of inertia $I'$ of an infinitely thin flux
tube of length $R$ is
\begin{equation}
I' (R) = {1\over 2} (A'R) R^2,
\end{equation}
with
\begin{equation}
A' = \sigma/6.
\end{equation}
The comparison of eq. (5.27) describing an infinitely thin flux tube with eq.
(5.25) gives a quantitative estimate of the increase of the moment of inertia
of the flux tube due to its finite thickness.

We now compare these results for $V_0 + V_{VD} = - L_I$ of the dual theory with
the "modified area law" model$^{[9]}$,
 eq. (2.22).  In the dual theory $i \log
W(\Gamma)$ is replaced by $i \log W_{eff}(\Gamma)$, given in the classical
approximation by eq.\ (5.1). 
  This gives in the limit of short
distances the perturbative expression eq. (3.7)
so that the short distance limit of the dual theory is the short range
component $ i \log W^{SR} (\Gamma)$.
The long distance limit of $i \log W_{eff} (\Gamma)$ is fixed by the values of
$\sigma$ and $A$.  Replacing $A$ by $A'$ in this limit yields $i \log W^{LR}
(\Gamma)$.  This shows that $i \log W^{LR} (\Gamma)$ describes a zero width
flux tube. Aside from this difference we see that
 the ``two components'' of eq. (2.22) arise
  as two limits of a single classical solution describing the evolution
 of the potential  produced by compression of the field lines 
with increasing $R$.

As the simplest example of the implications of $V_{VD}$, we add relativistic
kinetic energy terms to $-(V_0 + V_{VD})$ to obtain a classical Lagrangian, and
calculate classically the energy  and angular momentum of $q\bar q$ circular
orbits, which are those which have the largest angular momentum $J$ for a given
energy.  We find$^{[20]}$ a Regge trajectory $J$ as a function of $E^2$ which
for
large $E^2$ becomes linear with slope $\alpha' = J/E^2 = 1/8\sigma (1 -
A/\sigma)$.  Then (5.25) gives $\alpha' \approx 1/6.3\sigma$, which is close to
the
string model relation $\alpha' = {1\over 2\pi\sigma}$.  This comparison shows
how at the classical level a string model emerges when the velocity dependence
of the $q\bar q$ potential is included.  The fact that the difference between
the two expressions for $\alpha'$ is small indicates that the infinity narrow
string may be a good approximation to the finite width flux tube forms between
the $q\bar q$ pair.

To summarize:

1) The potential $V_0(R)$ is determined by eqs.(5.9) and (3.39) evaluated 
 at the static solution.

\noindent 2) The potential $V_{VD}$ is given by eqs. (5.22) and (3.40)
evaluated at the solution  of the classical equations to first order 
 in the velocity squared. The resultant integrals have been calculated 
 numerically$^{[1]}$ and determine four functions $V_{+}(R),
 V_{-}(R), V_{L}(R)$ and $V_{||}(R)$  which specify uniquely
 the terms  in the potential proportional to the velocity squared.
 Explicit expressions for these functions are given in reference 1.

\leftline{\bf Remarks}

  \noindent {\bf 1.}
 In the absence of quark sources $(G_{\mu\nu}^S = 0), {\cal
L}_{eff}$ describes a system of massive dual gluons and monopoles.  Because of
the dual Higgs mechanism there are no unwanted massless particles 
in the spectrum.  The massive
particles of the dual theory cannot be identified with the massive particles of
Yang--Mills theory, since the dual theory just describes the low energy
spectrum. These masses determine rather the scale $R_{FT}={1\over M}$
above which the dual theory should describe the $q\bar q$ interaction. Since a quark anti-quark pair moving in an orbit of
radius $R$ can only radiate
a particle of mass $M$ if ${1\over R} > M$, in the domain $R>{1\over M}$ where
the dual theory describes Yang--Mills theory no dual gluons or monopoles are
emitted. The glueballs of Yang--Mills theory, on the otherhand,
 are described by closed loops of
color flux, obtained by coupling the dual potentials to closed Dirac strings
and finding the corresponding static solution of the field equations of the
dual theory.

\noindent {\bf 2.}  The Lagrangian density ${\cal L}_{eff}$ (3.35) describes
the coupling of the Dirac string to Abelian configurations of dual potentials,
and the functional integral (3.41) for $W_{eff} (\Gamma)$ is restricted to such
configurations.  The external $q\bar q$ pair has in effect selected out a
particular sector of the dual theory relevant to the $q\bar q$ potential.  As a
consequence the resulting potential should not be very sensitive to the details
of the dual gauge group.

\noindent {\bf 3.}  The Dirac string in the classical solution was a straight
line connecting the
$q\bar q$ pair.  This gave the configuration having the minimum field energy
\footnote{The Dirac string of the dual theory, in contrast to that
 of electrodynamics, is physical. The vanishing of the Higgs field
on the string produces a vortex and an associated flux tube containing 
energy. This vortex can not be removed by a gauge transformation since
 such a transformation leaves the magnitude of the Higgs field 
 unchanged.}.
The flux tube corresponding to a given string position is concentrated in the
neighborhood of that string since the monopole current vanishes there.  To
evaluate the contributions to the potential arising from fluctuations of the
shape and length of the flux tube$^{[21]}$, 
we must integrate over field configurations
generated by all Dirac strings connecting the $q\bar q$ pair.  This
amounts to
doing a functional integral over all Dirac polarization tensors $G_{\mu\nu}^S
(x)$.  Similar integrals have recently been carried out by Akhmedov, et
al.,$^{[10]}$ in a somewhat different context.  The functional integral over
$G_{\mu\nu}^S (x)$ is replaced by a functional integral over corresponding
world sheets $y(\sigma,\tau)$, multiplied by an appropriate Jacobian. As a
result they obtain $^{[10]}$ an effective string theory free from the 
conformal  anomaly$^{[22]}$. 
 Such techniques when applied in the context of the
dual theory should lead to a corresponding effective string theory.

\section{The Classical Approximation for $V_{SD}$}

In this section we evaluate the expression for $V_{SD}$ given in Section four
using the classical solutions to the dual theory described in
Section five.  We consider separately the four contributions to $V_{SD}$ (See
eq. (\ref{vsd})):

\noindent {\bf (1)} $V_{Thomas}$:   Eq. (\ref{nvt})-(4.18)
 with
$V_0(R)$
determined by eq. (\ref{v0}).

\noindent {\bf (2)} $V_{LS}^{MAG}$:  Eq. (\ref{nvls}) with
$\langle \langle
\vec D \rangle\rangle_{eff}$ and $\langle\langle \vec H \rangle\rangle_{eff}$
replaced by their classical values $\vec D$ and $\vec H$, namely
\begin{equation}
V_{LS}^{MAG} = - \sum_{j=1}^2 {4\over 3} {e_j\over m_j} \vec S_j \cdot (\vec H
(\vec
z_j) - \vec v_j \times \vec D (\vec z_j)),
\label{e61}
\end{equation}
with $\vec H - \vec v_j \times \vec D$  calculated to first order in
the
velocity.  To this order the static field configurations follow the
motion of the quarks adiabatically and we find from eq. (3.37)
\begin{equation}
\vec H (\vec z_j) - \vec v_j \times \vec D (\vec z_j) = - \vec\nabla 
(C_0(\vec x) -
\vec C(\vec x) \cdot \vec v(\vec x))\Bigg|_{\vec x = \vec z_j},
\label{hq}
\end{equation}
where
\begin{equation}
\vec v(\vec x) = {\vec v_1 + \vec v_2\over 2} + \vec\omega \times \vec x,
\label{vel}
\end{equation}
and
\begin{equation}
\vec\omega = {\vec R \times {d\vec R\over dt}\over R^2} .
\label{omega}
\end{equation}
In eqs. (\ref{hq}-\ref{omega}), ${\vec v_1 + \vec v_2\over 2}$ is the
instantaneous velocity
of the origin of the coordinates which we have chosen as the midpoint of the
line $L$ connecting  the $q \bar q$ pair and $\vec\omega$ is the instantaneous
angular velocity of $L$.  (The motion of the $q \bar q$ pair along $L$ does not
contribute to eq. (\ref{hq})).

We can understand the result eq. (\ref{hq}), as
follows:  The left hand side is the color magnetic field at the position of the
$j$th quark in the Lorentz system in which it is instantaneously at rest.
The magnetic
field in this comoving system is determined by the gradient of the
corresponding dual scalar potential, namely $C_0 - \vec C \cdot \vec v$.  Indeed 
(6.2) remains valid beyond the classical approximation with the
replacement $C_0 \rightarrow \langle\!\langle C_0 \rangle\!\rangle_{eff}, \vec
C \rightarrow \langle\!\langle \vec C \rangle\!\rangle_{eff}$.

Choosing $\vec R$ to lie along the $z$ axis and using eqs. (5.21) and (5.10)
for $C_0$ and $\vec C$ we find:
\begin{equation}
C_0 - \vec C \cdot \vec v = \hat e_\phi \cdot {d\vec R\over dt} C_- (z,\rho),
\end{equation}
where $\rho, \phi, z$ are cylindrical coordinates, and
\begin{equation}
C_- (z,\rho) = C_-^D (z,\rho) + c_- (z,\rho),
\end{equation}
where
\begin{equation}
C_-^D (z,\rho) = {e \rho\over 4\pi R} \left\{{1\over\sqrt{\rho^2 +
(z-{R\over 2})^2}} - {1\over\sqrt{\rho^2 + (z + {R\over 2})^2}}\right\},
\label{cdm}
\end{equation}
and where $c_- (z,\rho)$ satisfies the equation
\begin{equation}
(\tilde\nabla^2 - 6g^2 B^2) c_- = 6g^2 B^2 C_-^D.
\label{dcm}
\end{equation}

The solution of the linear integral equation (\ref{dcm}) for $c_-$
determines, via eqs. (\ref{hq}) and (6.5) the non-perturbative part of the
color
magnetic field in the comoving Lorentz system.  From eqs.
(6.6)-(6.8) it follows that for any fixed value of $z$ and $\rho$ this field
vanishes like
${1\over R}$ for large $q\bar q$ separation.  The vanishing of this field at
large $R$ is in accordance with the observation of Buchmuller$^{[23]}$
that in a flux tube picture the color field in the comoving frame should be
purely electric.  However, for any finite value of the $q\bar q$ separation
there is a color magnetic field in this system, and eqs. (6.1)-(\ref{dcm}) give
\begin{equation}
\!\!\!\!V_{LS}^{MAG} = {V'_2 (R)\over R} \left\{\!\left({\vec S_1\!\cdot \!
(\vec R \times \vec p_1)\over m_1^2} - {\vec S_2\!\cdot \!(\vec R \times \vec
p_2)\over m_2^2}\right)\! + \!\left({\vec S_2 \!\cdot \!(\vec R \times \vec
p_1)\over m_1 m_2} - {\vec S_1 \!\cdot \!(\vec R \times \vec p_2)\over m_1 m_2}
\right)\!\!\right\}~~~~~~~~~~ ,
\end{equation}
where
\begin{equation}
V'_2 (R) = {4\over 3} \left\{{\alpha_s\over R^2} - {1\over 2\rho}
{\partial\over\partial \rho} [\rho c_- (\rho, z)]\Bigg|_{\rho = 0 \atop z
= R/2}\right\}.
\label{v2}
\end{equation}
The first term in (\ref{v2}) is the perturbative contribution to $V'_2 (R)$
arising from $C_-^D$ and the second term is the non-perturbative part which
behaves like ${1\over R}$ for large $R$ and which would not be present in the
simple flux tube picture of Buchmuller.

Finally adding $V_{LS}^{MAG}$ to
$V_{Thomas}$  gives the complete expression for the spin orbit
coupling $V_{LS}$,
\begin{equation}
V_{LS} = \!\left[{1\over R} {dV_0\over dR} +\! 2 {V'_1 (R)\over R}\right] \!\!
\left[\!{\vec S_1 \!\cdot \!\vec R \!\times \!\vec p_1\over 2m_1^2} - {\vec S_2
\!\cdot \!\vec R \!\times \!\vec p_2\over 2m_2^2} \right] \! + \!{V'_2 (R)\over
R} \! \left[{\vec S_2 \!\cdot \!\vec R \!\times \!\vec p_1\over m_1 m_2} -
{\vec S_1 \!\cdot\! \vec R \!\times \!\vec p_2\over m_1
m_2}\right],~~~~~~~~~~~~~~~~~~~~
\label{fvls}
\end{equation}
where
\begin{equation}
V'_1 (R) = V'_2(R) - {dV_0\over dR}.
\label{v1}
\end{equation}
Eq. (\ref{fvls}) expresses the spin orbit potential in terms of the central
potential and a single independent function $V'_2 (R)$ determined
by the dual scalar potential $C_0 - \vec C \cdot \vec v$ in the comoving frame.
This result for $V_{LS}$ satisfies identically the constraints
of Lorentz invariance (6.12) (The Gromes
Relations).$^{[24]}$ Furthermore, since $V'_2 (R) \rightarrow {1\over R}$ for
large
$R$, we have
\begin{equation}
\lim_{R\rightarrow\infty} V'_1 (R) = - {dV_0\over dR} = - \sigma,
\end{equation}
which is the value given by the flux tube model for all $R$.

\noindent {\bf (3)} $V_{SS}$:  Eq. (4.22) with ${\delta
\langle\!\langle \vec H (z_j)\rangle\!\rangle_{eff}\over\delta 
\vec H_{S} (z_{j^\prime} )}$
replaced by ${\delta \vec H (z_j)\over \delta \vec H_{S}
 (z_{j^\prime})}$.  Since, to
first
order in the velocity, $\vec C$ is determined by $\vec D_{S}$ alone (see eq.
(\ref{dc})) the ${\partial\vec C\over \partial t}$ term in $\vec H$ does not
contribute to its variational derivative with respect to $\vec H_{S}$ and eq.
(3.37) gives
\begin{equation}
{\delta  H_k (x)\over \delta H_{S\ell} (x^\prime)} =
\delta_{k\ell} \delta (\vec x - \vec x^\prime)
 \delta (t - t') - \nabla_k {\delta C_0
(x)\over\delta H_{S\ell} (x^\prime)}.
\label{e614}
\end{equation}
The quantity ${\delta C_0\over \delta \vec H_{S}}$ in turn satisfies the
equation obtained by taking the variational derivative of eq. (\ref{dc0}) with
respect to $\vec H_{S}$, namely
\begin{equation}
(\nabla^2 - 6g^2 B^2) {\delta C_0 (x)\over\delta H_{S\ell}
 ( x^\prime)}
 = \nabla_\ell \delta (\vec x - \vec x^\prime) \delta (t-t').
\label{dc0hs}
\end{equation}
The double integral in eq.(\ref{nnvss}) then becomes a
single integral over $t$ of the static quantity ${\delta \vec H (\vec
z_j)\over\delta \vec H_{S} (\vec z_{j^\prime})}$.
  We emphasize that this simplification
obtains only
in the classical approximation we are now considering.

Eqs. (\ref{e614}) and (\ref{dc0hs}) give
\begin{equation}
{\delta H_k (\vec x)\over\delta H_{S\ell} (\vec x^\prime)} 
= \delta_{k\ell}
\delta (\vec x - \vec x^\prime) + 
\nabla_k \nabla'_\ell G(\vec x, \vec x^\prime),
\label{e616}
\end{equation}
where the Green's function $G(\vec x, \vec x')$ satisfies
\begin{equation}
(-\nabla^2 + 6g^2 B^2 (\vec x)) G(\vec x, \vec x') = \delta (\vec x - \vec x').
\end{equation}
$G(\vec x, \vec x')$ is the potential at $\vec x$ due to a point charge at
$\vec x'$ in presence of the monopole charge density $j_0^{MON}$ (5.8) carried
by $B (\vec
x)$.  Since $B (\vec x)$ approaches its vacuum value $B_0$ as $\vec x
\rightarrow \infty, G$ vanishes exponentially at large distances, i.e.,
\begin{equation}
G(\vec x, \vec x')_{\overrightarrow{\vec x \rightarrow \infty}} - {e^{- m_B
|\vec x - \vec x'|}\over 4\pi |\vec x - \vec x'|},
\end{equation}
where
\begin{equation}
m_B^2 = 6g^2 B_0^2 = {6\pi\over\alpha_s} B_0^2 \approx {\pi\over 4}
{\sigma\over\alpha_s},
\end{equation}
and where we used the result, $\sigma \approx 24B_0^2$, obtained from the
energy per unit length of the static flux tube solution.  Using a value
$\alpha_s = .37$ obtained from fitting the $c\bar c$ and 
$b\bar b$ spectrum$^{[1]}$  we
obtain $m_B \approx 640 MeV$.

Separating off the Coulomb contribution to $G$ we have
\begin{equation}
G = -{1\over 4\pi |\vec x - \vec x'|} + G^{NP},
\label{e620}
\end{equation}
where $G^{NP}$ satisfies the equation
\begin{equation}
(-\nabla^2 + 6g^2 B^2) G^{NP} = - {6g^2 B^2 (\vec x)\over 4\pi|\vec x - \vec
x'|}.
\label{e621}
\end{equation}
Inserting eqs. (\ref{e616}) and (\ref{e620}) into eq. (4.22) gives
\begin{equation}
V_{SS} = V_{SS}^{spin} + V_{SS}^{1/m^2},
\end{equation}
where
\begin{equation}
V_{SS}^{spin} =  {4\over 3} {e^2\over m_1 m_2} \Bigg\{ (\vec S_1 \cdot
\vec S_2) \delta (\vec z_1 - \vec z_2) + {(\vec S_1 \cdot \vec\nabla) (\vec
S_2 \cdot \vec \nabla^\prime)}  G (\vec x, \vec x^\prime)
\Bigg|_{\vec x= \vec z_1, \vec x^\prime= \vec z_2 }\Bigg\},
\end{equation}
\begin{equation}
V_{SS}^{1/m^2} = -{4\over 3} \sum_{j=1}^2 {e^2\over 8m_j^2} \vec\nabla \cdot
\vec\nabla^\prime
 G^{NP} (\vec x, \vec  x^\prime)\Bigg|_{\vec x^\prime = 
\vec x = \vec z_j ,}.
\end{equation}
The potential $V_{SS}^{spin}$  is the same as previously obtained$^{[11]}$.
At small $R$ it approaches the usual
perturbative spin-spin interaction, and  at long
distances it is exponentially  damped due to screening by the monopole charge.
The spin independent contribution $V_{SS}^{1/m^2}$ of
$V_{SS}$ depends upon $R$ via the dependence in eq. (\ref{e621}) of $G^{NP}$ on
$B$. It was not included in ref.$^{[11]}$.

\noindent {\bf (4)} $V_{Darwin}$:  Eq.(4.24) with  $\langle\!\langle 
\hat G_{\mu\nu}
\rangle\!\rangle_{eff}$ replaced 
by $\hat G_{\mu\nu}$ , namely
\begin{equation}
\int V_{Darwin} dt = - {4\over 3} \sum_j {e\over 8m_j^2} \int_{\Gamma_{j}}
 dx^\mu
\partial^\nu \hat G_{\nu\mu} (x).
\label{nnvd}
\end{equation}
To evaluate (\ref{nnvd}) we note from eqs. (3.11) and (\ref{gmn2}) that
\begin{equation}
\partial^\nu \hat G_{\nu\mu} (x) = j_\mu (x),
\end{equation}
where $j_\mu (x)$ is the quark anti-quark current.  The monopole current
does not contribute to $\partial^\nu
\hat G_{\nu\mu}$ and  $V_{Darwin}$ becomes,
\begin{equation}
\int V_{Darwin} dt = - {4\over 3} \sum_j {e\over 8m_j^2} \int_{\Gamma_{j}}
dx^\mu j_\mu (x) 
= - {4\over 3} \sum_j {e_j\over 8m_j^2} \int dt \rho (z_j).
\label{fvd}
\end{equation}
Omitting self energy terms we insert $\rho (\vec z_1) = -  e\delta (\vec z_1 -
\vec z_2), \rho (\vec z_2) = e\delta (\vec z_2 - \vec z_1)$ into eq. (6.27) and
obtain
\begin{equation}
V_{Darwin} =  {e^2\over 6} \left({1\over m_1^2} + {1\over
m_2^2}\right) \delta (\vec z_1 - \vec z_2).
\end{equation}
In Appendix B we show that the alternate form (B.3) for $V_{Darwin}$ reduces in
the classical approximation to the same expression (6.28).

  There are  two then  spin independent terms proportional to $\left({1\over
m_1^2} + {1\over m_2^2}\right)$.  The first is $V_{SS}^{1/m^{2}}$  
(6.24). The second is  $V_{Darwin}$
(6.28). 

To summarize:
 In reference 4 the coefficient
of $\left({1\over m_1^2} + {1\over m_2^2}\right)$ in the velocity dependent
potential was written as:
\begin{equation}
V_{SS}^{1/m^2} + V_{Darwin} \equiv {1\over 8} \left({1\over m_1^2} + {1\over
m_2^2}\right)
\nabla^2 (V_0 (R) + V_a (R))
\end{equation}
which defines $V_a$.  Eqs. (6.24) and (6.28) give
\begin{equation}
\nabla^2 V_a = \nabla^2 V_0^{NP} (R) - {4\over 3}e^2 \vec \nabla \cdot \vec
\nabla^\prime  G^{NP} (\vec x, \vec x^\prime
)\Bigg \vert_{\vec x = \vec  x^\prime = \vec z_j},
\end{equation}
where $V_0^{NP} (R)$ is the non-perturbative part of the central potential so
that $V_a$ is determined by
the non-perturbative dynamics of Yang--Mills theory.  The first term in 
(6.30) is the color electric contribution to $V_a$ and the second is the color
magnetic contribution.

The spin dependent potential is then given by:
\begin{equation}
V_{SD} = V_{LS} + V_{SS}^{spin} + {1\over 8} \left({1\over m_1^2} + {1\over
m_2^2}\right) \nabla^2 (V_0 (R) + V_a (R))
\end{equation}
with $V_{LS}$ given by (6.10) and (6.11), $V_{SS}^{spin}$ by 
(6.23) and $V_a (R)$ by (6.30).  

It should be emphasized that these results along with those of section 5
 do not account for  quantum fluctuations about the classical solutions.  To
account for these fluctuations we must return to the more general eqs. 
(4.15), (4.22) and (4.24) and (4.3) which determine $V_{q\bar q}$ in the
dual theory independent of the classical approximation.

\section{Conclusion}

We have shown how the analysis of ref.$^{[3]}$
 of the heavy quark potential $V_{q\bar q}$ in terms
of Wilson loops $W(\Gamma)$ leads to the expression for the long
distance behavior of $V_{q\bar q}$ in terms of an effective Wilson
loop $W_{eff} (\Gamma)$ calculated in a dual theory describing a dual
superconductor.  The coupling of the dual theory to heavy quarks is then
uniquely specified with spin and relativistic effects accounted for
unambiguously to order $\left({1\over mass~quark}\right)^2$, the highest order
for which the concept of a potential makes sense.

The calculation of $W_{eff}(\Gamma)$ in the classical approximations leads to
expressions for the various terms in $V_{q\bar q}$ with clear physical
interpretations.  These results coincide for the most part with the expressions
for $V_{q\bar q}$ given by a previous dual theory calculation$^{[1]}$ in which
the $q\bar q$ motion was treated semi-classically.  The present treatment gives
an additional contribution, (6.29) to $V_{q\bar q}$, and the contribution
of ``Thomas precession'' now appears automatically whereas in the
semi-classical treatment it has to be put in by hand.

We can use the Wilson loop $W_{eff}
(\Gamma)$  to approximate
$W(\Gamma)$ for large loops, i.e.,
 for $R > {1\over M}$, where $M$ is   
either the mass of dual gluon $C_\mu$ or of the monopole field $B_i$
(about
 500MeV).
However, since the dual theory gives a heavy quark potential which 
approaches lowest order perturbation theory for small $R$, it should
remain applicable down to distances
where radiative corrections giving rise to a running coupling constant
become important.

Most significant is the fact that we have obtained
an expression for the $q\bar q$ potential in the dual theory which
makes no reference to the classical approximation . 
Furthermore since the formulae of  reference $^{[2]}$
are obtained 
 starting from a relativistic treatment
of the $q\bar q$ ~QCD interaction, the results provide a 
 direct connection of
the dual theory to QCD 
which could lead to an understanding of the constituent
quark model on a more fundamental level.

As a final remark we note that the dual theory we propose is an SU(3) gauge
theory, like the original Yang-Mills gauge theory.  However, the coupling to
quarks selected out only Abelian configurations of the dual potential.
Therefore, our results for the $q\bar q$ interaction do not depend upon the
details of the dual gauge group and should be regarded more as consequences of
the
general dual superconductor picture rather than of our particular realization
of it.  The essential feature of this picture is the description of long
distance Yang--Mills theory by a dual gauge theory in which all  particles
become massive via a dual Higgs mechanism.

\section*{Acknowledgements}

One of us (M.B) would like to thank M. Polikarpov and L. Yaffe for 
enlightening  conversations.

\section*{Appendix A}

Notice that
\renewcommand{\theequation}{\arabic{equation}}
\setcounter{equation}{0}

\begin{equation}
\vec a_j \equiv {4\over 3} {e_j\over m_j}  \langle\!\langle \vec D(\vec
z_j)\rangle\!\rangle_{eff}\tag{A1}
\end{equation}
can be interpreted as the acceleration of the $j$th quark so that eq. (4.17)
can be rewritten
\begin{equation}
V_{Thomas} = - {1\over 2} \sum_{j=1}^2 S_j \cdot (\vec v_j \times \vec a_j),
\tag{A2}
\end{equation}
which is the usual expression obtained from semiclassical considerations.  To
express $V_{Thomas}$ in terms of the derivative of the static potential we
first note from eq. (4.9) that
\begin{equation}
{\delta i\log W_{eff} (\Gamma)\over \delta \vec D_S (\vec x)} = {4\over 3}
\langle\!\langle \vec D (\vec x) \rangle\!\rangle_{eff}. \tag{A3}  
\end{equation}
Now, using the fact that
\begin{equation}
\vec\nabla_1 \vec D_S (\vec x) = - e{\bf 1} \delta 
(\vec x - \vec z_1), \tag{A4}
\end{equation}
where $\vec\nabla_1 = {\partial\over\partial\vec z_1}$, we have
\begin{equation}
\begin{array}{rl}
R {dV_0(R)\over dR} &= \vec R \cdot \vec\nabla_1 V_0\\
&= \vec R \int d\vec x {\delta i\log W_{eff}\over \delta\vec D_S (\vec x)}
\cdot \vec\nabla_1 \vec D_S(\vec x)\\
&= - e \vec R \cdot \int d\vec x {4\over 3} \langle\!\langle \vec D (\vec x)
\rangle\!\rangle_{eff} \delta (\vec x - \vec z_1)\\
&= - e {4\over 3} R \langle\!\langle \vec D(\vec z_1) \rangle\rangle_{eff}
\cdot \hat R.
\end{array} \tag{A5}
\end{equation}
Now by symmetry, $\langle\!\langle \vec D(\vec z_1)
\rangle\!\rangle_{eff}$ evaluated at the position of a quark must lie along
$\hat R$.  Hence,
\begin{equation}
- {4\over 3} e \langle\!\langle \vec D (\vec z_1) \rangle\!\rangle_{eff} = \hat
R{dV_0\over dR} . \tag{A6}
\end{equation}
Eq. (A1) then gives $\vec a_1 = - {\hat R\over m_1} {\partial V_0\over\partial
R}$, so that eq. (A2) gives (4.18).

\section*{Appendix B}

Here we begin with an alternate form for $V_{Darwin}$ where $A_\mu$ does not
appear explicitly.$^{[4]}$
\renewcommand{\theequation}{\arabic{equation}}
\setcounter{equation}{0}

\[
e \int dx^\mu \langle\!\langle D^\nu F_{\nu\mu}(x) \rangle\!\rangle =
\int_{t_{i}}^{t_{f}} dt \nabla^2 V_0 
\]

\[
-ie^2 \int_{\Gamma_j}
 dx^\mu\int_{\Gamma_j}
 dx'_\sigma [\langle\!\langle F_{\mu\nu} (x) F^{\sigma\nu}
(x') \rangle\!\rangle - \langle\!\langle F_{\mu\nu} (x) \rangle\!\rangle
\langle\!\langle F^{\sigma\nu} (x') \rangle\!\rangle]=
\]

\begin{equation}
=   \int_{t_{i}}^{t_{f}} dt \nabla^2 V_0 - ie^2 \int_{\Gamma_{j}} dt
\int_{\Gamma_{j}} dt' [\langle\!\langle
F_{0k}(z_j) F^{0k}(z_j^\prime)
 \rangle\!\rangle - \langle\!\langle F_{0k}(z_j) \rangle\!\rangle
\langle\!\langle F^{0k}(z_j^\prime) \rangle\!\rangle ]. \tag{B1}
\end{equation}
Eq.(4.12) and the relation between $\delta S^{\mu\nu}$ and variations of 
$G_{\mu\nu}^S$ give
\begin{equation}
e\int_{\Gamma_{j}} dt \int_{\Gamma_{j}} dt' {\delta\over 
\delta S^{0k}(z_j^\prime)}
\langle\langle F^{0k}(z_j) \rangle\rangle =  {4e^2\over 3} \int_{\Gamma_{j}} dt
\int_{\Gamma_{j}} dt' {\delta \langle\langle D_k (z_j) \rangle\rangle_{eff}\over
\delta D_{Sk} (z_j^\prime)} . \tag{B2}
\end{equation}
Then using eq.(2.21) with $z_2$ replaced by $z_1$ and e by -e we obtain
\begin{equation}
\int dt V_{Darwin} = \
\sum_j \int dt \nabla^2 V_0/8m_j^2 - {4e^2\over 3} \sum_j  \int_{\Gamma_{j}} dt
\int_{\Gamma_{j}} dt' {1\over 8m_j^2} {\delta \langle\!\langle D_k (z_j)
 \rangle\!\rangle_{eff}\over \delta D_{Sk} (z_j^\prime)}, \tag{B3}
\end{equation}
which gives a second form for $V_{Darwin}$.  The classical approximation to
(B3) is obtained by replacing
\[
{\delta \langle\!\langle \vec D(z_j)
 \rangle\!\rangle_{eff}\over \delta \vec
D_S (z_j^\prime)}\quad {\rm by} \quad
 {\delta\vec D (\vec z_j)\over \delta \vec D_S
(\vec
z_j)} \delta (t - t'). \]
This yields the expression
\begin{equation}
V_{Darwin} = \sum_{j=1}^2 \left[{1\over 8m_j^2} \nabla^2 V_0 (R) 
- {4\over 3}
{e^2\over 8m_j^2} {\delta D_k (\vec z_j)\over \delta D_{Sk} (\vec
z_j)}\right]. \tag{B4}
\end{equation}
Following the
same reasoning that led to eq. (A5) 
 we obtain
\begin{equation}
  {4\over 3} e^2
 {\delta  D_k (\vec z_1)\over \delta  D_{Sk} (\vec z_1)} =
 \nabla_1^2 V_0^{NP} (R) . \tag{B5}
\end{equation}
(There is no perturbative contribution to the left  hand side of (B5).)
The second term in  (B4) then cancels the non-perturbative part of
the first term.   Eq. (B4) then becomes
\begin{equation}
V_{Darwin} = \left({1\over 8m_1^2} + {1\over 8m_2^2}\right) \nabla^2 
\left( -{4\over 3} {e^2\over 4\pi R} \right) = {e^2} {\delta (\vec z_1 - \vec
z_2)\over 6} \left({1\over m_1^2} + {1\over m_2^2}\right) , \tag{B6}
\end{equation}
which coincides with (6.28).

\section*{References}
\begin{enumerate}

\item  M. Baker, James S. Ball and F. Zachariasen, Phys. Rev. {\bf D51} 1968
(1995).

\item  N. Brambilla, P. Consoli and G.M. Prosperi, Phys. Rev., {\bf D50} 5878
(1994).  

\item  A. Barchielli, N. Brambilla and G.M.
Prosperi, Nuovo Cimento {\bf 103A} (1990) 59.

\item  A. Barchielli, E. Montaldi and G.M. Prosperi, Nucl. Phys. {\bf B296}
(1988) 625; {\bf B303} 752(E) (1988).

\item  E. Eichten and F. Feinberg, Phys. Rev. {\bf D23} 2724 (1981).

\item  M.A. Peskin, in Proceedings of the 11th Stanford Linear Accelerator
Center Summer Institute (1983) SLAC report No. 207, ed. by P. McDonough, p. 151;
W. Lucha, F. Schoeberl and D. Gromes, Phys. Rep. {\bf 200} 127 (1991).

\item  N. Seiberg,  Nucl. Phys. {\bf B435} 129 (1995).

\item  N. Seiberg and E. Witten, Nucl. Phys. {\bf B431} 484 (1994).

\item  N. Brambilla and G.M. Prosperi, in 
``Quark Confinement and the Hadron Spectrum,'' ed. by N. Brambilla and G.M.
Prosperi, World Scientific Singapore, p. 195 (1995); 
Phys. Lett. {\bf B236}, 69 (1990)

\item  E.T. Akhmedov, M.N. Chernodub, M.I. Polykarpov and M.A. Zubkov, Phys. 
Rev. {\bf D53} 2097 (1996) (hep-th 9505070).

\item  M. Baker, J.S. Ball and F. Zachariasen, Phys. Rev. {\bf D44} 3949
(1991); Phys. Lett. {\bf 283} 360 (1992).

\item  For another dual theory, see S. Maedan and T. Suzuki, Prog. Theor. Phys.
{\bf 81} 229 (1989).

\item  S. Mandelstam, Phys. Rep. {\bf 23c} 245 (1976); G. t'Hooft, in ``Proc.
Eur. Phys. Soc. 1975,'' ed. by A. Zichichi (Ed. Comp. Bologna 1976) p. 1225.

\item  P.A.M. Dirac, Phys. Rev., {\bf 74} 817 (1948).

\item L. D. Landau and E. M. Lifshitz, Classical theory of fields (Pergamon,
 New York 1975), p. 149.

\item  G.G. Darwin, Phil. Mag. {\bf 39} 537 (1920).

\item  M. Baker, J.S. Ball and F. Zachariasen, Phys. Rev. {\bf D41} 2612
(1990).

\item  A.A. Abrikosov, Sov. Phys. JETP, {\bf 32},1442 (1957), H.B. Nielsen
 and P. Olesen, Nucl. Phys. {\bf B61} 45 (1973).

\item  M. Baker, J.S. Ball and F. Zachariasen, Phys. Rev. {\bf D44} 3328
(1991).

\item  M. Baker in ``Proceedings of the workshop on quantum infrared physics,''
Eds. H.M. Fried, B. Muller; World Scientific (1995), p.351.

\item  The leading correction to the static potential of $V_0(R)$ has the
universal value ${-\pi\over 12R}$.  [See M. Luscher, Nucl. Phys. {\bf B180} 317
(1981).]

\item  A.M. Polyakov, Phys. Lett.,{\bf 103B} 207, 211 (1981)

\item  W. Buchmuller, Phys. Lett., {\bf 112B} 749 (1982).

\item  D. Gromes, Z. Phys., {\bf C26} 401 (1984).
\end{enumerate}

\newpage

\begin{figure}
\centering
\epsfig{file=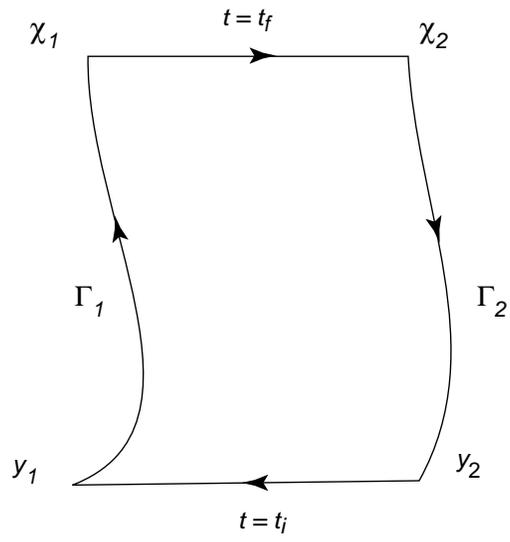, width=0.40 \textwidth}
\caption{Wilson loop for the quark anti-quark system.}
\end{figure}

\end{document}